\documentclass[12pt, a4paper]{article}
\usepackage[margin=1in]{geometry}

\usepackage{amsmath, amsfonts, amssymb, amsthm} 

\usepackage{graphicx, color}

\usepackage{multirow} 
\usepackage{makecell} 
\usepackage{float} 
\usepackage{todonotes}

\usepackage{tikz}
\usetikzlibrary{arrows.meta,shapes,shapes.arrows,
shapes.geometric,shapes.multipart,decorations.pathmorphing,
positioning,shapes.swigs}
\usepackage{enumerate, enumitem}
\setlist[itemize]{leftmargin=0.7cm, labelsep=0.5em, itemsep=0.0cm, topsep=0.1cm, partopsep=0cm, parsep=0cm} 
\setlist[enumerate]{leftmargin=1cm, labelsep=0.5em, itemsep=0cm, topsep=0cm, partopsep=0cm, parsep=0cm} 

\usepackage{natbib}
\usepackage{booktabs}
\usepackage{xurl, hyperref}
\hypersetup{
colorlinks=true,
linkcolor=blue,
filecolor=blue,
urlcolor=blue,
citecolor=blue
}

\usepackage{xr}
\makeatletter
\newcommand*{\addFileDependency}[1]{
\typeout{(#1)}
\@addtofilelist{#1}
\IfFileExists{#1}{}{\typeout{No file #1.}}
}
\makeatother
\newcommand*{\myexternaldocument}[1]{%
\externaldocument{#1}%
\addFileDependency{#1.tex}%
\addFileDependency{#1.aux}%
}
\myexternaldocument{New_SLDB_JRSSB_Supp}

\usepackage{algorithm} 
\usepackage{setspace} 
\usepackage{rotating} 

\usepackage{epstopdf} 
\usepackage{bm, bbm} 
\usepackage{titlesec} 
\usepackage{caption} 
\usepackage{mathtools} 
\usepackage{algpseudocode}
\graphicspath{ {plot/} }
\definecolor{mycolor}{RGB}{0,200,200} 
\definecolor{red1}{RGB}{255,64,64}
\definecolor{blue1}{RGB}{128,255,255}
\definecolor{green1}{RGB}{0,205,0}

\newtheorem{theorem}{Theorem}
\newtheorem{lemma}{Lemma}

\theoremstyle{definition}
\newtheorem{assumption}{Assumption}

\newtheorem{remark}{Remark} 

\makeatletter
\def\thm@space@setup{%
 \setlength{\thm@preskip}{5pt}
 \setlength{\thm@postskip}{5pt}
}
\makeatother
\newcommand{\indep}{\rotatebox[origin=c]{90}{$\models$} }

\DeclareMathOperator*{\argmin}{arg\,min}

\newcommand{\ind}{\mathbf{1}}

\usepackage{xspace}
\newcommand{\method}{SLDB\xspace}

\titlespacing*{\section}{0pt}{8pt}{6pt}
\titlespacing*{\subsection}{0pt}{8pt}{6pt}
\titlespacing*{\subsubsection}{0pt}{8pt}{6pt}

\titleformat{\section}
 {\normalfont\fontsize{15}{15}\bfseries}{\thesection}{1em}{}
\titleformat{\subsection}
{\normalfont\fontsize{13}{13}\bfseries}{\thesubsection}{1em}{}
\titleformat{\subsubsection}
 {\normalfont\fontsize{12}{12}\bfseries}{\thesubsubsection}{1em}{}

\usepackage{etoolbox}
\AtBeginDocument{ 
\everydisplay{\fontsize{12}{13}\selectfont}
}

\allowdisplaybreaks

\begin{document}

\setlength{\abovedisplayskip}{8pt}
\setlength{\belowdisplayskip}{8pt}
\setlength{\abovedisplayshortskip}{8pt}
\setlength{\belowdisplayshortskip}{8pt}
\setlength{\jot}{2pt}

\title{\vspace*{-2cm} \makebox[1cm][c]{Sliced $L^p$ Distributional Balancing}}
\author{
Haoran Zhang$^{a}$, Guanhua Chen$^{b}$, Chan Park$^{a}$\\
\makebox[1cm][c]{{\footnotesize $^{a}$Department of Statistics, University of Illinois Urbana-Champaign, Champaign, IL 61820, U.S.A.}}\\[-0.25cm]
\makebox[1cm][c]{{\footnotesize $^{b}$Department of Biostatistics and Medical Informatics, University of Wisconsin-Madison, Madison, WI 53706, U.S.A.}}
}
\date{}

\maketitle

\begin{abstract} 
A popular class of causal inference methods addresses confounding through weighting, which reweights treated and control groups to balance their covariate distributions without using outcome information, thereby preserving a design-based perspective. In this paper, we propose sliced $L^p$ distributional balancing (\method), a family indexed by $p\in[1,\infty)$ that measures imbalance by averaging squared $L^p$ distances between the cumulative distribution functions of one-dimensional linear projections. The Cram\'er--Wold device ensures that this criterion identifies equality of multivariate distributions, while projection reduces its computation to sorting-based one-dimensional operations. Because our method lies outside the maximum mean discrepancy (MMD) framework underlying many existing distributional balancing methods, their theoretical and computational tools do not directly apply. We therefore develop a computationally efficient projected subgradient descent algorithm for estimating balancing weights, offering improved computational complexity over MMD-based methods. Furthermore, we establish a novel theoretical framework for \method-based causal effect estimation and prove, under suitable conditions, $\sqrt{n}$-consistency and asymptotic normality, with the asymptotic variance attaining the semiparametric efficiency bound. Finally, we develop inferential procedures that do not require augmentation with an outcome model, thereby retaining the design-based principle. Simulation studies and a real-world application demonstrate that \method performs competitively with existing methods.

\end{abstract}

\newpage 

\section{Introduction} \label{sec:introduction} 

Confounding represents the fundamental challenge in causal inference based on observational studies. One widely adopted approach to addressing this challenge is the design-based framework, which seeks to reconstruct a study design resembling a randomised experiment without using outcome information. A well-constructed design provides a principled foundation for subsequent causal effect estimation. Notable examples of design-based methods include stratification, matching, and weighting \citep{Imbens2015}.

There is a long history of weighting methods spanning not only causal inference but also survey statistics and missing data analysis \citep{HorvitzThompson1952,Hajek1971,Deville1992,HiranoImbensRidder2003,Hainmueller2012,ImaiRatkovic2014,Zubizarreta2015,ChanYamZhang2016,athey2018approximate, WongChan2018,Zhao2019,Kallus2020,Hazlett2020,Wang2020,Hirshberg2021,ChenChenYu2023,Ertefaie2023, Kong2023, ChenHulingChenYu2024,Kim2024,HulingMak2024,de2025data,Shen2025,Kim2026,CFD2026}. In the terminology of causal inference, the primary objective of weighting is to assign an appropriate weight to each unit, thereby constructing a pseudo-population in which the covariate distributions of the treated and control groups are well balanced with respect to specified balancing criteria. Achieving such balance is expected to mitigate, and ideally eliminate, confounding, thereby enabling consistent estimation of the causal effects of interest. 

Existing balancing methods can be broadly categorised into three main strands according to how the balancing weights are constructed, as reviewed in Section~\ref{sec:Balancing review}: (a) by using the inverse propensity score \citep{Rosenbaum1983} as balancing weights; (b) by aligning a finite set of features or low-order moments across groups; and (c) by aligning the entire covariate distributions across groups, an approach we refer to as \textit{distributional balancing}. This paper contributes to the third strand by developing a new method for achieving distributional balance.

Within this third strand, most approaches have been formulated in terms of integral probability metrics (IPMs), with a prominent subclass based on maximum mean discrepancies (MMDs; \citealp{Gretton2012}) induced by reproducing kernel Hilbert spaces (RKHSes; \citealp{WongChan2018, ChenHulingChenYu2024, HulingMak2024, CFD2026}). Under suitable smoothness conditions, MMD-based methods can yield $\sqrt{n}$-consistent causal effect estimators. However, this statistical guarantee comes with practical and theoretical limitations. First, standard implementations involve dense kernel matrices, leading to substantial memory and computational costs as the sample size grows. Second, asymptotic normality generally requires augmenting the weighting estimator with an outcome regression and is not established for the unaugmented estimator.

Alternative distributional balancing approaches have considered non-MMD criteria. For example, Wasserstein balancing \citep{Kong2023, Yan2024} exhibits dimension-dependent convergence rates and, in general, does not yield $\sqrt{n}$-consistent causal effect estimators. More recently, \citet{de2025data} and \citet{Shen2025} developed forest kernel balancing, an outcome-adaptive MMD approach. Neither method establishes $\sqrt{n}$-consistency for the causal effect estimator, and by using outcome information to construct the forest kernel, both depart from a strictly outcome-free design.

Taken together, the existing literature lacks a computationally scalable, fully design-based distributional balancing method that delivers $\sqrt{n}$-consistent causal effect estimation with a characterised asymptotic distribution, without relying on outcome regression augmentation. These gaps naturally raise the following question. \\[-2.5em]
\begin{center}
\textit{Can we develop a distributional balancing method that \\
(a) retains the $\sqrt{n}$-consistency established for MMD-based distributional balancing;\\
(b) further characterises its asymptotic distribution;\\
(c) improves computational scalability; and \\
(d) preserves the design-based philosophy?}
\end{center}

To address the challenges discussed above, we develop a new design-based distributional balancing framework. Our main contributions are fivefold.

\begin{itemize}

\item[(a)] \textit{Methodology}. We propose \textit{sliced $L^p$
distributional balancing} (\method), which combines an $L^p$ discrepancy
between cumulative distribution functions with the Cram\'er--Wold device
\citep{cramer1936some}, which states that a multivariate distribution is
uniquely determined by its one-dimensional linear projections along all
directions. Motivated by this characterization, \method replaces direct
comparison of multidimensional covariate distributions with a collection of
one-dimensional comparisons. It measures the discrepancy along each projection
by the $L^p$ distance between the corresponding cumulative distribution
functions and averages the squared distances over projection directions; see
Section~\ref{sec:SWDBalancing} for details. This dimension reduction turns
multivariate distributional balancing into sorting-based one-dimensional
operations, substantially alleviating the statistical and computational
challenges associated with direct multidimensional distributional balancing. Notably, the underlying metric of \method is a non-MMD IPM except when $p=2$, in which case it reduces to an MMD, namely the energy distance. Consequently, existing MMD-based
theory cannot be directly applied to the general family, necessitating a new
theoretical analysis of the resulting weighting estimator.

\item[(b)] \textit{Theory}. In Section~\ref{sec:theory}, we develop a unified theory for every fixed $p\in[1,\infty)$. We establish a parametric rate for sliced $L^p$ balance and the $\sqrt n$-consistency of the proposed weighting estimator for the average treatment effect, matching the statistical guarantee
of competing MMD-based methods. Specifically, we derive a sufficient condition, formulated in terms of Sobolev smoothness of the outcome regression, under which the resulting causal effect estimator achieves the parametric convergence
rate. Although the resulting condition resembles those established for MMD-based balancing, its proof relies on fundamentally different techniques. Furthermore, under additional regularity conditions on the covariate law and
the inverse propensity score, we establish that the \method estimator is asymptotically normal, with asymptotic variance equal to the semiparametric efficiency bound, without ever fitting an outcome or propensity score model. While other estimators without outcome regression augmentation are known to be asymptotically normal \citep{HiranoImbensRidder2003, ChanYamZhang2016, Ertefaie2023}, they rely on either nonparametric estimation of the propensity score or calibration based on moment conditions, rather than on a distributional balancing approach. Thus, to the best of our knowledge, our result is the first to establish asymptotic normality for a \textit{distributional balancing} estimator without outcome regression augmentation.

\item[(c)] \textit{Computation}. We develop a projected (sub)gradient algorithm tailored to the non-MMD structure of the proposed balancing family; see Algorithm~\ref{alg:psd}. The algorithm exploits one-dimensional sorting, avoids constructing dense Gram matrices, and is naturally parallelisable across projection directions. We show that, under the stated regularity and tuning conditions, a per-iteration complexity of $\mathcal O(n^{2+\alpha})$, where $0<\alpha<1/(2d-1)$ and $d$ denotes the dimension of the covariates to be balanced, suffices to guarantee $\sqrt{n}$-consistency and asymptotic normality. This rate remains below the $\mathcal O(n^{3})$ complexity of generic quadratic programming implementations of MMD-based balancing. Our \method estimator also empirically exhibits faster computation than the competing nonparametric inverse probability weighting and outcome-augmented inverse probability weighting estimators considered in our numerical studies; see Sections~\ref{sec:simulation} and~\ref{sec:application}.

\item[(d)] \textit{Design-based Inference}. Because the balancing weights depend only on treatment and covariates, the same estimated design can be reused across multiple outcomes. Each additional outcome requires only outcome-specific effect estimation and the corresponding uncertainty quantification.  Consistent with this design-based separation between study design and outcome analysis, Section~\ref{sec:inference} develops inferential procedures that do not require outcome-regression modelling or augmentation. A closed-form plug-in variance estimator yields an asymptotically conservative but valid Wald-type confidence intervals without resampling, while subsampling \citep{Politis1994} remains valid under substantially weaker conditions. We also examine the standard nonparametric bootstrap \citep{Efron1994}, whose validity cannot be established for \method and which indeed exhibits undercoverage in our numerical studies.

\item[(e)] \textit{Extensions}. Finally, we
extend the \method framework beyond the average treatment effect to accommodate
a broad range of causal parameters and settings, including the local average
treatment effect, the average treatment effect on the treated, multicategory
treatments, and the learning of optimal treatment regimes; see Remark \ref{remark:extension}. Due to space
constraints, we present these additional results in
Appendix C.
 
\end{itemize}

The remainder of this paper is organised as follows.
Section~\ref{sec:preliminary} introduces the causal inference framework and
reviews existing balancing methods. Section~\ref{sec:methodology} presents the
proposed \method framework together with its scalable optimisation algorithm.
Section~\ref{sec:theory} establishes the asymptotic properties of the proposed
estimator and provides statistical inference procedures. Sections~\ref{sec:simulation} and~\ref{sec:application} evaluate the
proposed method through simulation studies and an empirical application,
respectively. Section~\ref{sec:discussion} provides some concluding remarks.

\section{Preliminary}\label{sec:preliminary}

\subsection{Setup}\label{sec:Setup}

Let $\{(X_i, A_i, Y_i)\}_{i=1}^n$ be $n$ independent and identically distributed (i.i.d.) observations from a population. For each unit $i$, $X_i \in \mathcal{X} \subset \mathbb{R}^d$ represents a $d$-dimensional vector of pre-treatment covariates, $A_i \in \{0, 1\}$ is the treatment indicator such that $A_i = 1$ indicates unit $i$ receives treatment, and $Y_i \in \mathbb{R}$ is the observed outcome. Note that the outcome $Y$ may be of any type, such as binary, polytomous, or continuous. For $a \in \{0,1\}$, let $e_{a}(x)\equiv P(A = a\mid X = x)$ denote the propensity score \citep{Rosenbaum1983}. Let $n_1$ and $n_0$ denote the number of treated and control units, respectively, i.e., $\sum_{i=1}^{n} A_i = n_1$ and $n_0=n-n_1$. We adopt the potential outcomes framework \citep{Rubin1974} throughout the paper. For each unit $i$, let $Y_i(1)$ and $Y_i(0)$ denote the potential outcomes under treatment and control, respectively. We omit the subscript $i$ unless necessary.

Let $F_X$ denote the cumulative distribution function (CDF) of the marginal covariate distribution, and let $F_{wX\mid A=a}$ denote the corresponding weighted CDF among units with treatment status $A=a$ under the non-negative weight function $w(X,A)$. Similarly, let $F_n$ denote the empirical CDF of the covariates in the full sample, and let $F_{n,a,w}$ denote the weighted empirical CDF among units with treatment status $A=a$ induced by the non-negative weight vector $w=(w_1,\ldots,w_n)^\top \in \mathbb{R}^{n}$. Specifically, letting $\ind(\cdot)$ denote the indicator function, we have:
\begin{align*}
 &
 F_{X}(t) = \mathbb{E} \{ \ind (X \le t) \} \ , \quad 
 &&
F_{wX|A=a}(t) = 
\frac{ \mathbb{E} \{ w(X,A) \ind ( X \le t) | A=a \} }{ \mathbb{E} \{ w(X,A) | A=a \} }
\ ,
 \\
 &
 F_n(t) = \frac{1}{n} \sum_{i=1}^n \ind(X_i \le t) \ , \quad 
 &&
F_{n, a, w}(t) = 
\frac{ \sum_{i : A_i = a} w_i \ind(X_i \le t) } { \sum_{i : A_i = a} w_i }
\ , \quad 
&&
a \in \{0,1\} \ .
\end{align*} 

We impose the following condition on the covariate support.
\begin{assumption}[Covariate Support] \label{assumption:cpt}
The covariate support is $\mathcal{X}=[0,1]^d$.
\end{assumption}
\noindent
This assumption can be relaxed to more general compact and bounded covariate spaces with Lipschitz boundaries. We maintain the unit cube assumption to simplify the exposition and avoid unnecessary technicalities. Similar assumptions on the covariate support are standard in the distributional balancing literature \citep{ChanYamZhang2016,WongChan2018,CFD2026}.

We also introduce the following notation. We use $\indep$ to denote statistical independence between random variables. For a sequence of random variables $\{T_n\}$ and a sequence of positive constants $\{a_n\}$, we write $T_n = \mathcal{O}_p(a_n)$ and $T_n = o_p(a_n)$ if $T_n/a_n$ is bounded in probability and converges to zero in probability, respectively, as $n \rightarrow \infty$. For sequences of positive constants $\{a_n\}$ and $\{b_n\}$, we write $a_n \lesssim b_n$ if there exists a constant $C>0$ such that $a_n \leq Cb_n$ for all sufficiently large $n$, and $a_n \asymp b_n$ if $a_n \lesssim b_n$ and $b_n \lesssim a_n$. We also write $a_n\propto b_n$ if there exists a constant $C>0$ such that $a_n=Cb_n$ for all $n$. Let $\xrightarrow{p}$ and $\xrightarrow{d}$ denote convergence in probability and convergence in distribution, respectively. Finally, let $\mathcal{W}^{s,2}(\mathcal{X})$ denote the fractional Sobolev space of functions on $\mathcal{X}$.

The primary estimand considered in this paper is the average treatment effect (ATE), denoted by $\tau = \mathbb{E}\{ Y(1) - Y(0)\}$. While \method is applicable to a broader class of causal estimands, we use the ATE as a running example for exposition. Extensions to other causal effects are presented in Appendix C. In order to establish identification of the ATE based on observational data, we make the following standard causal inference assumptions; see \citet{Imbens2015} and \citet{Hernan2020} for textbook discussions.
 
\begin{assumption}[Causal Inference] \label{assumption-causal}
(a) $Y = Y(A)$ almost surely; (b) $Y(a) \perp\!\!\!\perp A | X$ for $a \in \{0,1\}$; (c) there exists a constant $\eta > 0$ such that $e_{1}(x) \in [\eta, 1-\eta]$ for all $x \in \mathcal{X}$.
\end{assumption}

\noindent
Under Assumption \ref{assumption-causal}, the ATE is identified by 
\begin{align}
 \tau
 =
 \mathbb{E}\{ \mu_1(X) -\mu_0(X)\} 
 = 
 \mathbb{E} \bigg[
 \bigg\{ \frac{A}{e_1(X)}
 - \frac{1-A}{e_0(X)}
 \bigg\} Y
 \bigg] \ ,
 \label{eq:IPW}
\end{align}
where $\mu_a(X) \equiv \mathbb{E}(Y\mid A=a,X)$ is the outcome 
regression function for treatment group $a \in \{0, 1\}$. 

\subsection{Balancing Weights and Existing Approaches} \label{sec:Balancing review}

The second representation of \eqref{eq:IPW} motivates weighting estimators for the ATE of the form
\begin{equation}
\widehat{\tau}_{\text{weight}} = \sum_{i=1}^n w_i (2A_i-1) Y_i \ , 
\qquad w=(w_1,\ldots,w_n)^\top \in \Omega \ ,
\label{eq:5}
\end{equation}
where $\Omega \subset \mathbb{R}^n$ denotes the admissible set of weights. A standard choice is
\begin{align}
 \Omega \equiv \bigg\{ w \in \mathbb{R}^n : w_i\geq 0 \text{ for all $i \in \{1,\ldots,n\}$}, \sum_{i : A_i = 1} w_i =\sum_{i : A_i = 0} w_i = 1 \bigg\} \ .
 \label{eq:Omega}
\end{align} 
These constraints make $\sum_{i:A_i=a}w_i\delta_{O_i}$ a probability measure for each $a\in\{0,1\}$, where $\delta_v$ is the Dirac measure at $v$, and interpret $w_i$ as unit $i$'s relative contribution within its treatment group. 

This weighting estimator admits a natural design-based interpretation. Specifically, the weights define a weighted pseudo-population in which treatment assignment is effectively independent of the observed covariates, so the ATE can be estimated by a difference in weighted outcomes without outcome modelling. This requires each weighted treatment-group covariate distribution to match the target-population distribution:
\begin{align}
 F_{wX|A=1} = F_{X} = F_{wX|A=0} . 
 \label{eq-balancing criteria}
\end{align}
Inverse propensity weights $w(X,A)\propto 1/e_A(X)$ achieve this condition; see Lemma~\ref{lemma:oracle}. 

Motivated by this criterion, balancing methods seek to construct weights such that the corresponding empirical distributions are well aligned, namely $F_{n,1,w} \simeq F_{n} \simeq F_{n,0,w}$. Intuitively, the closer these empirical distributions are, the more faithfully the weighted sample approximates the ideal pseudo-population, and the more reliable the resulting weighting estimator \eqref{eq:5} is expected to be. 

Existing methods differ in how they estimate or approximate the oracle inverse propensity weights. IPW uses the estimated propensity score in the Horvitz--Thompson estimator or its normalised H\'ajek analogue \citep{HorvitzThompson1952,Hajek1971}, with the propensity score estimated using either parametric or nonparametric methods. Parametric IPW can be easily implemented using standard regression techniques; however, its validity relies on correct specification of the propensity score model.  In contrast, nonparametric IPW relaxes this requirement by estimating the propensity score over a flexible function class. For instance, \citet{HiranoImbensRidder2003} and \citet{Ertefaie2023} show that sieve IPW and highly adaptive lasso (HAL) IPW estimators, respectively, are asymptotically linear for the ATE under their respective regularity conditions. These approaches reduce reliance on parametric specification, although their implementation may require careful tuning in practice to ensure that their theoretical properties hold.

Alternatively, moment-based methods aim to achieve \eqref{eq-balancing criteria} by directly balancing a set of test functions $\phi(X)$. Examples include the covariate balancing propensity score (CBPS; \citealp{ImaiRatkovic2014}), entropy balancing \citep{Hainmueller2012}, and their variants \citep{Zubizarreta2015,ChanYamZhang2016,Wang2020,ChenChenYu2023}. The effectiveness of these methods depends critically on the choice of $\phi$; specifically, if the selected test functions are insufficient to eliminate confounding bias, the resulting estimators may perform poorly.

Motivated by these limitations, recent work has focused on distributional balancing methods that directly target distributional alignment in \eqref{eq-balancing criteria}. 
Let $\mathcal{D}$ denote a discrepancy measure between two probability distributions such that $\mathcal{D}(\mu,\nu)=0$ if and only if $\mu=\nu$. The distributional discrepancies among the three distributions in \eqref{eq-balancing criteria} can then be jointly minimised as:
\begin{align} \label{eq-D balancing} 
 \min_{w} \left\{ \mathcal{D}( F_{wX|A=1} , F_{X} ) + \mathcal{D}(F_{wX|A=0} , F_{X}) + \mathcal{D}(F_{wX|A=1} , F_{wX|A=0} )
 \right\} ,
\end{align}
where attaining a minimum value of zero is equivalent to satisfying \eqref{eq-balancing criteria}. The first two terms align each weighted treatment group with the target
covariate distribution, and the third directly aligns the two weighted
treatment groups. 

Most existing distributional balancing methods relies on IPMs, taking $\mathcal{D}$ in \eqref{eq-D balancing} to be
\begin{align} \label{eq:IPM}
 \text{IPM}_{\mathcal{F}}(\mu, \nu) = \sup_{f \in \mathcal{F}} \mathbb{E}\{ f(V)-f(W) \} \ , \quad V \sim \mu, \ W \sim \nu,
\end{align}
where $\mathcal{F}$ is a prespecified critic class that determines the estimator's statistical and computational properties.

Taking $\mathcal{F}$ as the unit ball of an RKHS yields the MMD \citep{Gretton2012}, encompassing kernel balancing \citep{ChenHulingChenYu2024}, energy distance balancing \citep{HulingMak2024}, characteristic function distance balancing \citep{CFD2026}, and related functional balancing methods \citep{WongChan2018}. For important choices including energy distance and Mat\'ern-type kernels, the induced RKHS is a (possibly homogeneous) Sobolev space, expressing kernel assumptions through familiar smoothness conditions. Although this structure facilitates theoretical analysis and finite-dimensional optimisation, existing results for unaugmented estimators establish only $\sqrt{n}$-consistency and do not characterise the limiting distribution, including whether asymptotic normality holds. Asymptotic normality has been established only with outcome regression augmentation. In terms of implementation, MMD-based methods generally involve solving quadratic programs that require storing and factorizing a dense $n\times n$ Gram matrix, resulting in $\mathcal{O}(n^2)$ memory and often $\mathcal{O}(n^3)$ computational costs; see Section~\ref{subsec:relation}. Scalable kernel balancing \citep{Kim2024} addresses this computational bottleneck through low-rank approximation, but its theoretical properties remain to be fully understood.

Forest kernel balancing, proposed independently by \citet{de2025data} and \citet{Shen2025}, instead learns a data-adaptive and outcome-dependent kernel induced by ensemble trees. These methods can capture nonlinearities and tree-structured interactions, thereby reflecting a geometry distinct from that of Sobolev RKHSs. However, using outcomes to learn the kernel precludes a fully design-based approach and necessitates sample splitting and cross-fitting. Moreover, neither approach establishes $\sqrt{n}$-consistency or asymptotic normality for the unaugmented estimator, and whether the distributional balance in \eqref{eq-balancing criteria} is asymptotically attained remains unclear.

Beyond RKHSs, \citet{Kong2023} and \citet{Yan2024} use the class of multivariate 1-Lipschitz functions as the critic function class. By the Kantorovich--Rubinstein duality theorem, the corresponding IPM is the 1-Wasserstein distance \citep{villani2003topics}. While this approach admits a natural interpretation through optimal transport theory, Wasserstein balancing suffers from the curse of dimensionality and generally fails to yield $\sqrt{n}$-consistent estimation. Thus, moving beyond MMD introduces statistical and computational challenges.

\section{Methodology}\label{sec:methodology}
\subsection{The Sliced \texorpdfstring{$L^p$}{Lp} Distributional Balancing Framework}
\label{sec:SWDBalancing}

To address the aforementioned limitations of existing distributional balancing approaches, our construction combines two ideas. First, distributional balancing is particularly easy in one dimension, where it reduces to comparing one-dimensional CDFs. Such comparisons are computationally tractable and permit efficient computation of distributional discrepancies via sorting, without constructing or factorizing a dense Gram matrix. Second, the Cram\'er--Wold device---a multivariate distribution is uniquely determined by the collection of its one-dimensional linear projections \citep{cramer1936some}---provides a way to extend such one-dimensional comparisons to multivariate distributions. Together, these ideas motivate defining a multivariate discrepancy through one-dimensional projections. 

For ease of exposition, we introduce additional notation. Let $\mathbb S^{d-1}=\{\theta\in\mathbb R^d:\|\theta\|_2=1\}$, and let
$\sigma$ be the uniform probability measure on $\mathbb S^{d-1}$. For a probability measure $\mu$ on $\mathbb R^d$, write $\theta_\#\mu$ for its push-forward under the map $v\mapsto\theta^\top v$, and let $F_\mu^\theta$ denote the CDF of $\theta_\#\mu$. For any fixed $p\in[1,\infty)$, we define the \emph{sliced $L^p$ CDF discrepancy} as
\begin{align}
 \rho_p(\mu,\nu)
 &\equiv
 \int_{\mathbb S^{d-1}}
 \left\{\int_{\mathbb R}
 |F_\mu^\theta(t)-F_\nu^\theta(t)|^p\,dt\right\}^{2/p}
 d\sigma(\theta).
 \label{eq:rho-p}
\end{align}
For each direction, the inner integral in \eqref{eq:rho-p} determines the $L^p$ norm of the difference between the two projected CDFs. The discrepancy $\rho_p$ squares this one-dimensional $L^p$ discrepancy and averages it over all projection directions on the unit sphere. Thus, $\rho_p$ is a sliced squared $L^p$ CDF distance, where $p$ controls how differences between the projected CDFs are aggregated within each projection. This projection-and-average construction has been widely used across disciplines \citep{Rabin2011,Bonnotte2013,Bonneel2015,Nadjahi2021}. By the Cram\'er--Wold device, $\rho_p(\mu,\nu)=0$ if and only if $\mu=\nu$. Therefore, $\rho_p$ provides a criterion for distributional balance because it characterises alignment of the full multivariate distributions.

It is instructive to consider two particular choices of $p$. For $p=1$, we have
$\rho_1(\mu,\nu)=\int_{\mathbb S^{d-1}} W_1^2(\theta_\#\mu,\theta_\#\nu),d\sigma(\theta)$, which follows directly from the well-known representation of the 1-Wasserstein distance between one-dimensional distributions, given by $W_1(\mu,\nu)=\int_{\mathbb R}|F_{\mu}(t)-F_{\nu}(t)| dt$. For $p=2$, $\rho_2$ coincides with the energy distance \citep{ED2004} up to a multiplicative constant; we verify this identity in Appendix B. More generally, in the same Appendix, we show that $\rho_p$ is a squared IPM for every $p\in[1,\infty)$, while it admits a squared MMD representation if and only if $p=2$. Thus, our method of distributional balancing based on $\rho_p$ is an IPM-based framework that is generally not MMD-based, with $p=2$ as the sole exception.

We now use $\rho_p$ to turn the population balance condition
\eqref{eq-balancing criteria} into an objective. For a fixed $p \in [1,\infty)$, define
\begin{align}
 \mathcal L_{\mathrm{pop},p}(w)
 \equiv{}& \rho_p(F_{wX\mid A=1},F_X)
 +\rho_p(F_{wX\mid A=0},F_X)
 +\rho_p(F_{wX\mid A=1},F_{wX\mid A=0}).
 \label{eq:Lpop-p}
\end{align}
Although these terms quantify imbalance differently as
$p$ varies, they are minimised at the same population oracle weights for all $p$. The following lemma formalizes this point.
\begin{lemma}\label{lemma:oracle}
For every $p\in[1,\infty)$, (a) $\mathcal L_{\mathrm{pop},p}(w)=0$ if and
only if
$F_{wX\mid A=1}=F_X=F_{wX\mid A=0}$; and (b) the normalized oracle inverse
propensity score weights
$\widetilde w(x,a)=P(A=a)/e_a(x)$ achieve exact distributional balance,
so $\mathcal L_{\mathrm{pop},p}(\widetilde w)=0$.
\end{lemma}
\noindent 
Lemma~\ref{lemma:oracle} shows that changing $p$ does not alter the definition of exact balance or the oracle weights that achieve it; it changes only how the criterion measures departures from exact balance. The resulting family therefore provides a common framework for sliced distributional balancing. Importantly, for $p\neq2$, $\rho_p$ does not admit a squared MMD representation, so the RKHS-based optimisation and theoretical tools underlying existing MMD-based balancing methods do not directly apply. This distinction motivates the new estimation strategy and theoretical analysis. At the same time, the one-dimensional structure induced by projection allows the empirical objective to be evaluated through sorting-based operations, providing a basis for computationally efficient optimisation. We develop these aspects in the remainder of the paper.

\subsection{Estimation} \label{subsec:estimation}
 
To construct a sample criterion for a fixed $p\in[1,\infty)$, we approximate the spherical integral in \eqref{eq:rho-p} using randomly sampled projection directions. Specifically, we independently draw $L_n$ directions $\theta_1,\ldots,\theta_{L_n}$ from $\sigma$ and replace the spherical integral with its empirical average. Define
\begin{align}
 \widehat\rho_p(\mu,\nu)
 \equiv \frac{1}{L_n}\sum_{\ell=1}^{L_n}
 \left\{\int_{\mathbb R}
 \left|F_\mu^{\theta_\ell}(t)-F_\nu^{\theta_\ell}(t)\right|^pdt
 \right\}^{2/p}.
 \label{eq:rhohat-p}
\end{align}
We suppress the dependence of $\widehat\rho_p$ on $L_n$ and the sampled directions for notational simplicity. Replacing each $\rho_p$ term in the population criterion \eqref{eq:Lpop-p} with its sample counterpart $\widehat\rho_p$ and adding a ridge penalty yields the regularised sample objective:
\begin{align}
 \widehat{\mathcal{J}}_p(w) \equiv{}&
 \widehat\rho_p(F_{n,1,w},F_n)
 +\widehat\rho_p(F_{n,0,w},F_n)
 +\widehat\rho_p(F_{n,1,w},F_{n,0,w})
 +\lambda_n \|w\|_2^2,
 \label{eq:hatJ}\\
 \widehat w_p \equiv{}&
 \argmin_{w\in\Omega}\widehat{\mathcal{J}}_p(w).
 \label{eq:hatw}
\end{align}
The number of projections $L_n$ and the regularisation parameter $\lambda_n$ are tuning parameters, whose rates for the theoretical analysis are specified in Section~\ref{sec:theory}. From an optimisation perspective, the ridge penalty makes $\widehat{\mathcal{J}}_p$ strongly convex and stabilises the resulting weights. More importantly, it is essential for establishing the statistical validity of the resulting ATE estimator; see Theorem \ref{Theorem:asymptotics}.

The slicing-based construction also leads to a computationally convenient sample objective. Once the directions are sampled, each projected empirical CDF is a one-dimensional step function. Hence, for any $p\in[1,\infty)$, the inner integral in \eqref{eq:rhohat-p} reduces to a finite weighted sum over the intervals determined by the sorted projected observations. Consequently, $\widehat{\mathcal{J}}_p(w)$ can be evaluated using only one-dimensional sorting and arithmetic operations for every member of the sliced $L^p$ family.

For optimisation, $\widehat{\mathcal{J}}_p$ is everywhere differentiable for $p=2$ and differentiable almost everywhere for $p \ne 2$, allowing both cases to be handled within a unified projected (sub)gradient algorithm; see Algorithm~\ref{alg:psd}. Throughout, we use the convention that $\partial\widehat{\mathcal{J}}_p$ denotes the subdifferential when $p\neq 2$ and the standard gradient when $p=2$. The projection step maps each iterate onto the feasible set $\Omega$ in \eqref{eq:Omega}, thereby enforcing the weight constraints. Under the standard diminishing step-size conditions $\sum_t\eta_t=\infty$ and $\sum_t\eta_t^2<\infty$, the iterates converge to the unique minimiser $\widehat w_p$ \citep{Boyd2003}. 

\begin{algorithm}[!htp]
\caption{Projected (Sub)gradient Descent for Sliced $L^p$ Balancing Weights}
\label{alg:psd}
\footnotesize
\begin{algorithmic}
\State \textbf{Inputs} $\{(X_i, A_i)\}_{i=1}^n$, $p$, $\{\theta_\ell\}_{\ell=1}^{L_n}$, step sizes $\{\eta_t\}$, maximum number of iterations $T_{\max}$
\State \textbf{Initialization} Set $t=0$ and $w^{(0)} \in \Omega$ with $w_j^{(0)} = 1/n_{A_j}$
\Repeat
 \State Compute $g_p^{(t)} \in \partial \widehat{\mathcal{J}}_p(w^{(t)})$; see Appendix B for an expression of $\partial \widehat{J}_p$
 \State $w^{(t+1)} \leftarrow \text{P}_{\Omega} (w^{(t)} - \eta_t g_p^{(t)})$, where $\text{P}_{\Omega}$ is the projection onto $\Omega$ in \eqref{eq:Omega}
 \State $t \leftarrow t + 1$
\Until{convergence or $t = T_{\max}$}
\State \textbf{Output} Optimized weights $\widehat w_p = w^{(t)}$
\end{algorithmic}
\end{algorithm}

While Algorithm \ref{alg:psd} provides a general procedure for estimating $\widehat{w}_p$, its implementation requires tuning several hyperparameters, including the regularisation parameter $\lambda_n$, the number of random projections $L_n$, and the step size $\eta_t$, as well as specifying convergence criteria. We provide a ``default'' data-driven, yet design-based, procedure for selecting these hyperparameters and convergence criteria without using outcome information. Further implementation details are provided in Appendix A.

\subsection{Computational and Theoretical Implications of Slicing} \label{subsec:relation}
 
The first key advantage of \method over MMD-based balancing is computational scalability. Existing MMD-based distributional balancing methods, such as those of \cite{HulingMak2024,CFD2026,ChenHulingChenYu2024}, are typically formulated as constrained quadratic programs. Solving these programs generally requires constructing and storing dense Gram matrices, resulting in $\mathcal{O}(n^2)$ memory requirements and approximately $\mathcal{O}(n^3)$ computational complexity with generic solvers \citep{Nocedal2006}. A similar computational bottleneck arises for \citet{WongChan2018}. Although their formulation does not fall within the standard MMD framework, its eigenvalue optimisation has the same memory and time complexities. Moreover, because these optimisation problems are coupled through the Gram matrix, they offer limited opportunities for efficient parallelisation.

By contrast, \method exploits one-dimensional projections to obtain a more scalable optimisation
procedure. Projecting and sorting the observations costs
$\mathcal O(L_n n\log n)$ once, after which an objective or (sub)gradient
evaluation costs $\mathcal O(L_n n)$ per iteration; storing all projected
orderings requires $\mathcal O(L_n n)$ memory \citep{Rabin2011}; see
Appendix A for details. In Section~\ref{sec:theory}, we show that choosing $\alpha=0$ and $L_n\propto n$ suffices for
$\sqrt n$-consistency, giving $\mathcal O(n^2)$ work per iteration, below the cubic complexity of generic MMD-based solvers.
Even for asymptotic normality, it suffices to take $L_n\propto n^{1+\alpha}$ for any $0<\alpha<1/(2d-1)$, yielding a per-iteration complexity of $\mathcal{O}(n^{2+\alpha})$, which remains substantially below the cubic rate. We note that under $L_n\propto n^{1+\alpha}$ the $\mathcal{O}(L_n n)$ memory for cached orderings becomes $\mathcal{O}(n^{2+\alpha})$, slightly exceeding the $\mathcal{O}(n^2)$ storage of MMD-based solvers. This is easily avoided by re-sorting the projections on the fly, which lowers the memory requirement to $\mathcal{O}(n)$ while raising the per-iteration cost to $\mathcal{O}(L_n n\log n)$, still below cubic. Furthermore, computations across projection directions are independent and can therefore be parallelised with essentially no communication overhead. These features make \method particularly well suited to large-scale settings in which MMD-based distributional balancing becomes computationally demanding. We demonstrate these computational advantages empirically in Sections~\ref{sec:simulation} and~\ref{sec:application}.
 
The second key contribution is theoretical. Slicing permits parametric-rate inference even for distributional balancing methods based on non-MMD IPMs. Direct multivariate $W_1$ balancing (e.g., \citet{Kong2023}) uses multivariate 1-Lipschitz functions as the critic function class and typically exhibits the dimension-dependent empirical convergence rate $n^{-1/d}$ \citep{VanHandel2014}, even under arbitrarily smooth outcome regressions. This slow convergence can leave residual imbalance too large for $\sqrt n$-consistent treatment effect estimation \citep{CFD2026}. For $p=1$, $\rho_1$ instead averages squared $W_1$ distances between one-dimensional projections. These one-dimensional discrepancies admit parametric convergence rates under sufficient smoothness of the outcome regression; see Theorem~\ref{Theorem:asymptotics}. However, because $\rho_p$ does not in general admit an MMD representation for $p\neq2$, these guarantees cannot be readily obtained from existing RKHS-based theory. The next Section establishes them for every fixed $p\in[1,\infty)$, thereby extending distributional balancing theory beyond the MMD framework.

\section{Theory}\label{sec:theory}

\subsection{$\sqrt{n}$-consistency and Asymptotic Normality} \label{sec:consistency and normality}
Throughout this Section, we fix $p\in[1,\infty)$. We establish theoretical guarantees for the \method weights $\widehat w_p$ in \eqref{eq:hatw} and for the corresponding ATE estimator
\begin{equation}
 \widehat\tau_p
 \equiv \sum_{i=1}^n \widehat w_{p,i}(2A_i-1)Y_i.
 \label{eq:tau-p}
\end{equation} 

\begin{assumption}[Hyperparameters] \label{assumption:alg para}
In \eqref{eq:hatJ}, the regularisation parameter satisfies
$\lambda_n\asymp n^{-\alpha}$ for some constant
$0\leq\alpha<1/(2d-1)$. Furthermore, the number of random projections satisfies $L_n \ge \frac{12(4d)^{1/p}}{\kappa} \frac{n}{\lambda_n}$ for some constant $\kappa \in (0,1)$.
\end{assumption} 

\noindent
The first condition governs the decay rate of the regularisation sequence $\lambda_n$. Setting $\alpha = 0$ corresponds to a fixed regularisation parameter $\lambda_n = \lambda$, which is already sufficient to guarantee the $\sqrt{n}$-consistency of the resulting ATE estimator. On the other hand, establishing asymptotic normality requires eliminating the asymptotic bias, which necessitates a vanishing ridge penalty and thus requires $\alpha > 0$ (i.e., undersmoothing); see Theorem \ref{Theorem:asymptotics}.

The second condition regarding $L_n$ ensures that the Monte Carlo approximation error from the random projections does not dominate the statistical error, thereby preserving the theoretical convergence rates of the ATE estimator. Its order, $L_n\gtrsim n/\lambda_n\asymp n^{1+\alpha}$, is common across $p$, while the multiplicative constant depends on $p$. 

While these conditions provide asymptotic guidance for choosing $\lambda_n$ and $L_n$ as $n\to\infty$, in practice, we observe only a single finite sample of size $n$. We therefore provide a practical, data-driven procedure for tuning these hyperparameters, which we use in both our simulations and data application; see Appendix A for details.

\begin{assumption}[Outcome Regression Smoothness] \label{assumption:smoothness}
For each $a\in\{0,1\}$, the outcome regression function satisfies
$\mu_a\in\mathcal W^{s,2}(\mathcal X)$ for some $s>d/2+\max\{1/p,1/2\}$.
\end{assumption}

\noindent Assumption~\ref{assumption:smoothness} imposes a smoothness condition on the outcome regression, with the required degree of smoothness depending on $p$. Similar conditions are common in the distributional balancing literature \citep{Kallus2020,WongChan2018,Hirshberg2021}, and our requirement is nearly minimax-optimal. According to \citet{Robins2008}, obtaining $\sqrt{n}$-consistent ATE estimation without restricting the propensity score fundamentally requires the outcome regression to satisfy $\beta_\mu > d/2$\footnote{\label{fn:embedding} \citet{Robins2008} established this result in terms of H\"older smoothness. Under Assumption \ref{assumption:cpt}, H\"older smoothness can be translated into the corresponding Sobolev smoothness condition. Specifically, for any bounded compact domain $\mathcal{X}$ and any $\epsilon>0$, we have $C^\beta(\mathcal{X}) \subset W^{\beta-\epsilon,2}(\mathcal{X})$ where $C^{\beta}(\mathcal{X})$ denotes the class of functions with H\"older smoothness $\beta$ on $\mathcal{X}$. Consequently, if $\beta_{\mu}>d/2$, then $\mu_a\in \mathcal{W}^{s,2}(\mathcal{X})$ for some $s>d/2$.}. In the most conservative case, our condition requires $s > d/2 + 1$, exceeding this baseline by at most one derivative; see Appendix B for the technical origin of this gap. 
To position Assumption~\ref{assumption:smoothness} among existing balancing methods, let $\mathcal{H}_G$ and $\mathcal{H}_{\mathrm{ED}}$ denote the RKHSes associated with the Gaussian kernel \citep{WongChan2018} and the energy distance kernel\footnote{$\mathcal{H}_{\text{ED}}$ is a homogeneous Sobolev class without the boundedness of $\mathcal{X}$, but under Assumption \ref{assumption:cpt}, $\mathcal{H}_{\text{ED}}$ becomes a usual Sobolev class.} \citep{HulingMak2024}, respectively. These spaces satisfy the inclusion chain:
\begin{equation}\label{eq:sobolev_chain_main}
    \mathcal{H}_{G} \subset \bigcap_{s>0} \mathcal{W}^{s,2}(\mathcal{X}) \subset \mathcal{W}^{d/2+\max\{1/p,1/2\},2}(\mathcal{X}) \subset \mathcal{W}^{(d+1)/2,2}(\mathcal{X}) \subset \mathcal{H}_{\mathrm{ED}} \subset \mathcal{W}^{d/2,2}(\mathcal{X}).
\end{equation}
Consequently, our estimator accommodates a substantially broader function class than Gaussian kernel balancing while requiring only marginally stronger regularity than energy balancing.

\begin{assumption}[Outcome Moment Conditions] \label{assumption:outcome var}
For $a \in \{0, 1\}$, the conditional variance 
$\sigma_a^2(x) \equiv \mathrm{Var}( Y\mid A=a, X = x )$ is uniformly bounded 
over $x \in \mathcal{X}$, and $\mathrm{Var}\{\mu_a(X)\} < \infty$.
\end{assumption}
\noindent
Assumption \ref{assumption:outcome var} collects standard regularity conditions on the outcome distribution that bound the stochastic noise in the estimator \citep{HulingMak2024, CFD2026}. Note that this condition is automatically satisfied when assuming $Y$ is bounded.

While Assumptions~\ref{assumption:cpt}-\ref{assumption:outcome var} suffice to establish $\sqrt{n}$-consistency of the ATE estimator, two additional regularity conditions are needed to establish its asymptotic normality:

\begin{assumption}[Bounded Covariate Density] \label{assumption:density} 
The marginal distribution of $X$ admits a density $f_X(x)$ on the $d$-dimensional unit cube $\mathcal{X}$ satisfying $0 < \underline{f} \le f_X(x) \le \overline{f} < \infty$.
\end{assumption}

\begin{assumption}[Inverse Propensity Score Smoothness] \label{assumption:propensity_smoothness}
For each $a\in\{0,1\}$, the inverse propensity score satisfies
$1/e_a\in\mathcal W^{s,2}(\mathcal X)$ for some $s>d/2+\max\{1/p,1/2\}$.
\end{assumption}

\noindent Assumption \ref{assumption:density} is a standard regularity condition ensuring that the distribution of $X$ is sufficiently well-spread over its support $\mathcal{X}$. Assumption \ref{assumption:propensity_smoothness} requires the propensity score to possess a level of smoothness compatible with that imposed on the outcome regression. This smoothness condition plays a key role in recovering the oracle inverse propensity score weights and establishing asymptotic normality; see the discussion following Theorem \ref{Theorem:asymptotics}. Design-based weighting estimators generally require an analogous smoothness condition on the propensity score to achieve asymptotic normality. For instance, \citet{HiranoImbensRidder2003} require the propensity score $e_a$ to be continuously differentiable of order $\beta > 7d$; \citet{ChanYamZhang2016} require its differentiability $\beta>13d$;  \citet{Ertefaie2023} require the true propensity score to have finite sectional variation norm. Given footnote \ref{fn:embedding}, the smoothness condition of Assumption \ref{assumption:propensity_smoothness} is strictly weaker than that of \citet{HiranoImbensRidder2003} and \citet{ChanYamZhang2016}. By contrast, the finite sectional variation norm condition of \citet{Ertefaie2023} is neither weaker nor stronger than our assumption; see Appendix B for clarifying this point.

We now state the main result of the paper.

\begin{theorem}\label{Theorem:asymptotics}
Fix $p\in[1,\infty)$ and suppose that Assumptions \ref{assumption:cpt}-\ref{assumption:outcome var} hold. Note that $\alpha$ denotes the exponent of the regularisation parameter $\lambda_n \asymp n^{-\alpha}$.
\begin{enumerate}
    \item[(a)] The weights $\widehat w_p$
    achieve sliced $L^p$ balance at rate $\mathcal O_p(\lambda_n/n)$
    and induce squared bias of the same order, i.e., for $a \in \{0,1\}$, we have:
    \begin{align*}
        \rho_p(F_{n,a,\widehat w_p},F_n)
        &=\mathcal O_p\left(\frac{\lambda_n}{n}\right),
        \qquad 
        \left\{\int_{\mathbb R^d}\mu_a(x)\,
        d(F_{n,a,\widehat w_p}-F_n)(x)\right\}^2
        =\mathcal O_p\left(\frac{\lambda_n}{n}\right)
    \end{align*}
    if either
    (i) $\alpha=0$; or (ii) $0<\alpha<1/(2d-1)$ and
    Assumption \ref{assumption:density} additionally holds.
    \item[(b)] Under either case (i) or (ii) above, the \method ATE estimator is $\sqrt n$-consistent, i.e.,
    $(\widehat\tau_p-\tau)^2=\mathcal O_p(n^{-1})$.
    \item[(c)] Suppose $0<\alpha<1/(2d-1)$ and Assumptions \ref{assumption:density} and \ref{assumption:propensity_smoothness} additionally hold. Then, the weights $\widehat w_p$ recover the oracle inverse
    propensity score in the empirical $\ell^2$ norm:
    \begin{align*}
        \frac{1}{n}\sum_{i=1}^n\mathbf{1}(A_i=a)
        \left\{n\widehat w_{p,i}-\frac{1}{e_a(X_i)}\right\}^2
        =o_p(1),\qquad a\in\{0,1\}.
    \end{align*}
    Moreover, the \method ATE estimator is asymptotically normal as $\sqrt n(\widehat\tau_p-\tau)\xrightarrow{d}        \mathcal N(0,\sigma_{\text{ATE}}^2)$, where $\sigma_{\text{ATE}}^2$ equals the semiparametric efficiency
    bound for the ATE in the nonparametric model \citep{Hahn1998}:
    \begin{align*}
        \sigma_{\text{ATE}}^2=\mathbb E\left[\frac{\sigma_1^2(X)}{e_1(X)}
        +\frac{\sigma_0^2(X)}{e_0(X)}
        +\{\mu_1(X)-\mu_0(X)-\tau\}^2\right].
    \end{align*}
\end{enumerate}

\end{theorem}

\noindent Theorem \ref{Theorem:asymptotics} establishes three main results. First, the \method weights constitute a \textit{bona fide} set of balancing weights, with both the distributional discrepancy measured by $\rho_p$ and the squared bias induced by covariate imbalance converging to zero at rate $\mathcal{O}_p(\lambda_n/n)$. Second, this bias control yields $\sqrt{n}$-consistency under two distinct regularisation regimes. When $\alpha=0$, corresponding to a fixed ridge penalty, $\widehat\tau_p$ is $\sqrt{n}$-consistent under the baseline assumptions, without requiring the covariate distribution to admit a density or the propensity score to satisfy additional smoothness conditions. When $0<\alpha<1/(2d-1)$, the penalty vanishes with $n$, removing the asymptotic shrinkage induced by regularisation; in this regime, the additional density assumption is needed to control the empirical distributional approximation underlying the sliced balancing criterion.

The third result establishes recovery of the oracle inverse propensity weights in the empirical $\ell^2$ norm and, building on this recovery, asymptotic normality of the \method estimator without outcome regression augmentation. Moreover, the resulting asymptotic variance $\sigma_{\text{ATE}}^2$ coincides with the semiparametric efficiency bound for the ATE in the nonparametric model  \citep{Hahn1998}. Asymptotic normality without outcome regression augmentation has previously been established for several inverse probability weighting and entropy balancing estimators \citep{HiranoImbensRidder2003, ChanYamZhang2016, Wang2020, Ertefaie2023}. By contrast, existing asymptotic normality results for distributional balancing estimators rely on outcome regression augmentation \citep{WongChan2018,Hirshberg2021}. Part (c) of Theorem \ref{Theorem:asymptotics} therefore provides a genuinely new result; to the best of our knowledge, it is the first to establish asymptotic normality for a distributional balancing estimator without outcome regression augmentation.

How does \method attain this result without outcome regression augmentation? The asymptotic normality argument requires the bias induced by regularisation to be $o_p(n^{-1/2})$ and the estimated weights to recover the oracle inverse propensity weights in the empirical $\ell^2$ norm. Assumptions \ref{assumption:alg para} and \ref{assumption:propensity_smoothness} provide the key primitive conditions for establishing these two requirements.

First, by part (a) of Theorem \ref{Theorem:asymptotics}, asymptotic normality requires $\lambda_n\to 0$ (i.e., undersmoothing), so that the bias becomes $o_p(n^{-1/2})$. At the same time, $\lambda_n$ cannot vanish too quickly, as sufficient regularisation must be retained to control the estimated weights and facilitate recovery of the oracle inverse propensity weights. The rate restriction $\lambda_n\asymp n^{-\alpha}$ with $0<\alpha<1/(2d-1)$ reflects these competing requirements. The importance of such rate-dependent tuning is familiar from nonparametric and machine learning estimation, where tuning parameters govern the relative magnitudes of bias and stochastic error and, consequently, the attainable convergence rate and limiting distribution.

Second, Assumption \ref{assumption:propensity_smoothness} provides the link between distributional balance and recovery of the oracle inverse propensity weights. Intuitively, the \method criterion controls imbalance over a Sobolev class (see Appendix B for details). Thus, when the inverse propensity score belongs to this class and $\lambda_n$ does not vanish too quickly, such control translates into recovery of the oracle inverse propensity weights in the empirical $\ell^2$ norm.
 
\begin{remark}[Extensions to other causal estimands] \label{remark:extension}
While the theoretical guarantees in Theorem~\ref{Theorem:asymptotics} are presented for the ATE, our \method framework naturally extends to a broader class of causal estimands, including the average treatment effect on the treated (ATT), local average treatment effects (LATE), subgroup effects, multicategory treatments, and optimal treatment regimes. We establish the corresponding theoretical properties in Appendix C.
\end{remark}

\subsection{Design-based Inference} \label{sec:inference}

For each fixed $p$, we consider three design-based approaches to construct confidence intervals (CIs) for $\tau$: (a) Wald CIs based on a plug-in variance estimator, (b) subsampling CIs, and (c) bootstrap CIs. Corresponding hypothesis tests can be obtained by inverting these CIs. Section~\ref{sec:simulation} reports simulation evidence comparing the three approaches.

\subsubsection{Wald-Type Inference}
\label{sec:inference-wald}

Although Theorem~\ref{Theorem:asymptotics} establishes asymptotic normality
for $\widehat\tau_p$, the asymptotic variance $\sigma_{\text{ATE}}^2$ is
unknown. For a scalar constant $c \in \mathbb R$, consider the
plug-in variance estimator
\begin{align} \label{eq-plugin variance-c}
 \widehat{\sigma}_{\text{ATE},p}^2(c)
 =\frac{1}{n}\sum_{i=1}^n
 \left\{n\widehat w_{p,i}(2A_i-1)(Y_i-c)-\widehat\tau_p\right\}^2 .
\end{align}
 A notable advantage of this variance estimator is its computational efficiency, as it requires only the calculation of an empirical variance. Moreover, the estimator is theoretically valid in the sense that it converges to a value no smaller than the asymptotic variance. The following theorem formalizes this result.

\begin{theorem}
\label{thm:variance-c-main}
Fix $p \in [1,\infty)$, and suppose that Assumptions~\ref{assumption:cpt}-\ref{assumption:propensity_smoothness}
hold.
\begin{enumerate}
\item[(a)] For every fixed $c \in \mathbb R$, we have 
\begin{align*} 
\widehat{\sigma}_{\text{ATE},p}^2(c) \;\xrightarrow{p}\; \sigma^2_{\text{ATE}} + D(c) \ , \
 D(c) \equiv \mathbb{E}\!\left[ \frac{\big\{ e_0(X) \mu_1(X) + e_1(X)\mu_0(X) - c\big\}^2}{e_1(X)e_0(X)} \right] \geq 0 . 
\end{align*}
\item[(b)] Moreover, $D(c)$ is uniquely minimised at
\begin{align*}
 c^* =\frac{\mathbb{E} \left\{ \mu_1(X)/ e_1(X) +  \mu_0(X)/ e_0(X)\right\}}
 {\mathbb{E} \left\{ 1/ e_1(X) +  1/ e_0(X)\right\}}. 
\end{align*}
\item[(c)] Let $\widehat{c} = {\sum_{i=1}^{n} \widehat{w}_{p,i}^2 Y_i } / 
 { \sum_{i=1}^{n} \widehat{w}_{p,i}^2} $. Then, $\widehat{c} \xrightarrow{p} c^*$ and $\widehat{\sigma}^2_{\text{ATE},p}(\widehat{c})\xrightarrow{p}\sigma^2_{\text{ATE}}+D(c^*)$. 

\end{enumerate}
\end{theorem} 
\noindent 
Theorem~\ref{thm:variance-c-main} shows that the plug-in variance estimator computed on the raw outcome ($c=0$) is asymptotically conservative with the gap $D(0) \ge 0$. Centring the outcome by an appropriately chosen constant can shrink this gap, and the optimal choice $c^*$ can be consistently estimated from the observed data.  Nevertheless, Wald CIs based on $\widehat{\sigma}_{\text{ATE}}^2(\widehat{c})$ may remain asymptotically conservative, depending on the underlying data-generating law. Specifically, unless $e_0\mu_1+e_1\mu_0$ is constant, the optimal gap $D(c^*)$ remains strictly positive. Roughly speaking, greater variation in the outcome regressions and/or propensity scores across $X$---reflecting stronger confounding---can lead to a greater degree of conservatism. This conservatism represents an inherent limitation of the plug-in variance estimator that cannot be eliminated through optimal centring.

In principle, $\sigma_{\text{ATE}}^2$ can be consistently estimated by additionally estimating the outcome regressions and incorporating them into the variance estimator. Nevertheless, we advocate the conservative, design-based variance estimator in \eqref{eq-plugin variance-c} because of its practical advantages. In particular, $\widehat{\sigma}_{\text{ATE},p}^2(\widehat{c})$ requires minimal additional computation, essentially amounting to the calculation of an empirical variance, and is straightforward to implement and interpret. Moreover, our simulation studies and empirical application in the following Sections indicate that the resulting conservatism is generally modest in practice, further supporting the practical appeal of this approach.

\subsubsection{Subsampling}
\label{sec:inference-subsampling}

Subsampling \citep{Politis1994, Politis1999} approximates the sampling
distribution of $\widehat\tau_p$ by re-estimating it on subsamples of size
$m<n$ drawn without replacement. For each fixed $p$, a known convergence
rate and a nondegenerate limiting distribution suffice for asymptotically
exact CIs whenever $m\to\infty$ and $m/n\to0$. In particular, under
Theorem~\ref{Theorem:asymptotics}, subsampling asymptotically removes the
conservativeness of the plug-in Wald CIs.

These theoretical guarantees involve two practical considerations. First, the subsample size $m$ must be selected in finite samples. We adopt data-driven approaches implemented in existing software, such as the \texttt{moonboot} R package \citep{moonboot}, with further details provided in Appendix A. Second, unlike the closed-form Wald estimator, subsampling requires repeated estimation using resampled data. Nevertheless, our simulations confirm that this additional computational overhead remains modest.

\subsubsection{Bootstrap}
\label{sec:inference-bootstrap}

The standard nonparametric bootstrap \citep{Efron1994} approximates the sampling distribution of $\widehat\tau_p$ by repeatedly estimating $\widehat w_p$ on resamples of size $n$ drawn with replacement. Its validity depends on the regularity of the resulting estimator; roughly speaking, the estimator must behave smoothly under local perturbations of the data. However, our \method estimation procedure involves empirical sorting and optimisation over the simplex, making it nontrivial to formally establish the required regularity conditions. Therefore, we do not establish bootstrap validity for the \method estimator and urge caution when applying the bootstrap for any choice of $p$.

Consistent with these concerns, the simulation results in Section~\ref{sec:simulation} show that the bootstrap can undercover, mirroring the findings of \citet{CFD2026} for MMD-based estimators. Moreover, repeatedly solving the full optimisation problem is substantially more computationally expensive than either the Wald calculation or subsampling, as also demonstrated in our simulation study. Given its lack of established theoretical guarantees and substantially greater computational cost, we do not recommend the bootstrap-based inference for the \method estimator.

\begin{remark}[Multiple Outcomes] Beyond the inferential guarantees above, the design-based nature of \method offers a practical advantage in studies with multiple outcomes. Because the estimated balancing weights depend only on treatment and covariates, they can be computed once and reused for all outcomes measured on the same units. For Wald inference, each additional outcome therefore requires only the corresponding effect estimate and empirical variance. A similar computational saving applies to subsampling: the weights need to be estimated only once within each subsample and can then be reused across outcomes. By contrast, augmented IPW (AIPW, \citealp{Robins1994, Chernozhukov2018}) requires a separate outcome regression for each outcome and, under cross-fitting, repeated estimation of these regressions across folds.
\end{remark}

\section{Simulation}\label{sec:simulation}
\subsection{Simulation Design} 
We conduct a simulation study to  evaluate the finite-sample performance of point estimation and statistical inference across $n \in \{1000, 2000, 4000\}$ observations. The data-generating process (DGP) is configured with $d = 10$ pre-treatment covariates generated independently from a uniform distribution over $\mathcal{X} = [0,1]^{10}$, i.e., $X = (X_1, \ldots, X_{10})^\top \overset{\text{i.i.d.}}{\sim} \mathrm{Uniform}(\mathcal{X})$. To introduce nonlinear confounding that affects both treatment assignment and potential outcomes, we define $C(X) = (X_3 - 0.5)^2 - (X_4 - 0.5)^2$, a centred quadratic term that captures nonlinear curvature. The treatment assignment indicator $A \in \{0, 1\}$ is generated from a Bernoulli distribution with propensity score 
\begin{align*}
 e(X)
 = \text{expit}
 \bigg\{ 0.2 - 0.3 X_1 + 0.3 X_2 + 6 C(X) + 0.05 \sum_{j=5}^{10} (X_j - 0.5)
 \bigg\} . 
\end{align*}

The potential outcomes are generated as $Y(a) = \mu_a(X) + \varepsilon_a$ for $a = 0,1$, where
\begin{align*}
&\mu_0(X) = 0.4 X_1 + 0.4 X_2 + 0.2 \sum_{j=5}^{10} X_j,
&&\mu_1(X) = \mu_0(X) + 1.5 C(X),
\end{align*}
and the errors $\varepsilon_0$ and $\varepsilon_1$ are independently drawn from the standard normal distribution. Under this DGP, the ATE is $0$, and the plug-in variance estimator in \eqref{eq-plugin variance-c} is conservative. 

For our proposed method, we focus on two representative estimators with prominent geometric interpretations: $p = 1$ (SL1DB), corresponding to the sliced squared 1-Wasserstein distance, and $p = 2$ (SL2DB), corresponding to the energy distance. We compare these estimators with four classes of competing methods. First, we consider parametric approaches, including inverse probability weighting (IPW) and covariate balancing propensity scores (CBPS). For both methods, the propensity score is estimated using a main-effects logistic regression model and is therefore potentially subject to model misspecification. Second, we consider two nonparametric methods outside the distributional balancing framework: HAL-IPW \citep{Ertefaie2023} and AIPW implemented using double machine learning \citep{Chernozhukov2018}. Third, we consider MMD-based distributional balancing: Gaussian kernel balancing (MMD-G) and energy balancing (MMD-EB), which rely on quadratic programming, and scalable kernel balancing (MMD-SKB) \citep{Kim2024}, which uses a low-rank spectral approximation of the Gram matrix to reduce this cost but provides neither asymptotic theory nor an inferential procedure for the ATE estimator.

For statistical inference, the available procedures differ across estimators. For our \method estimators, we construct nominal 95\% CIs using the three strategies detailed in Section~\ref{sec:inference}: the closed-form Wald CI based on the plug-in standard error (PSE) from \eqref{eq-plugin variance-c}, subsampling CI (SS), and the bootstrap CI (Boot). For the parametric estimators (IPW and CBPS) and MMD-SKB, we use the standard nonparametric bootstrap; we remark that the bootstrap CI for MMD-SKB is included only for comparison, as its theoretical validity for this estimator has not been established. For the other MMD-based methods (MMD-G and MMD-EB), we use subsampling. For HAL-IPW and AIPW, which are asymptotically normal and semiparametrically efficient, we construct Wald CIs by plugging the estimated propensity score and outcome regression into the efficient influence function and using its empirical variance. For all subsampling and bootstrap procedures, we use 500 resamples. We evaluate the empirical performance of these inference procedures in terms of coverage probability and average CI length. Complete implementation details for all estimators and inference procedures are provided in Appendix A.

We execute $1000$ independent Monte Carlo replications for each sample size $n \in \{1000, 2000, 4000\}$. Point estimation is evaluated using empirical bias and root mean squared error (RMSE) against the true ATE ($\tau = 0$). We also compare the RMSE with the semiparametric efficiency bound (SEB) to assess finite-sample performance relative to this benchmark, which represents the fundamental efficiency limit for ATE estimators under nonparametric approaches.

\subsection{Results}
Table~\ref{tab:sim_point_estimation} summarises the performance of the competing estimators and their associated inferential procedures. For point estimation, the parametric methods (IPW and CBPS) exhibit a persistent bias of approximately $0.04$ at all sample sizes, accompanied by noticeably inflated RMSEs. These results illustrate the sensitivity of parametric approaches to model misspecification in the presence of nonlinear confounding and distributional discrepancies. In contrast, the nonparametric IPW methods (HAL-IPW and AIPW) and distributional balancing methods (MMD-G, MMD-EB, MMD-SKB, SLDB) reduce empirical bias to negligible levels at $N=4000$ and achieve uniformly lower and comparable RMSEs, although SKB exhibits nontrivial bias at $N=1000$ and $2000$.

Comparing the RMSE with the SEB further reveals the finite-sample efficiency of the competing estimators relative to the theoretical efficiency limit. The ratios of RMSE to SEB are close to one for HAL-IPW, AIPW, SL1DB, and SL2DB, consistent with the theoretical result that all four estimators asymptotically attain the semiparametric efficiency bound. HAL-IPW and AIPW, which additionally exploit outcome information, enjoy only a marginal finite-sample advantage, suggesting that the outcome-agnostic \method estimators cause no substantial efficiency loss.

For inference, we first focus on the \method estimators, for which we advocate the Wald and subsampling CIs in Section \ref{sec:inference}. Consistent with our theoretical results, both CIs appear to attain nominal coverage. Although the proposed plug-in variance estimator in \eqref{eq-plugin variance-c} is theoretically conservative for this DGP, our simulation results suggest that the degree of conservatism is mild, with $\mathrm{PSE}/\mathrm{SEB}$ remaining around $1.05$ across the evaluated settings. Comparing the two approaches, the Wald CIs are even slightly shorter than the subsampling CIs. Consistent with the RMSE comparison, the Wald CIs of HAL-IPW and AIPW also attain near nominal coverage and are marginally shorter than those of SL1DB and SL2DB. Taken together, these findings suggest that the proposed Wald procedure offers a practically attractive approach to inference, combining reliable coverage and modest conservatism with substantially faster computation than the resampling-based procedures used by the other distributional balancing methods; see Table \ref{tab:sim_time}. In contrast, the standard bootstrap exhibits mild to severe undercoverage across the considered specifications and is therefore fundamentally unreliable, at least in the context of this simulation study.

For the other methods, we find that the subsampling CIs attain nominal coverage, with interval lengths that are practically indistinguishable across the methods using subsampling (MMD-G, MMD-EB, and \method). By contrast, bootstrap inference is generally invalid for these methods, including MMD-SKB, consistent with the findings of \citet{CFD2026}. Finally, we note that IPW and CBPS also exhibit bootstrap undercoverage, which appears to be driven by their estimation bias.

\begin{table}[!htp]
\centering
\scriptsize
\setlength{\tabcolsep}{2.5pt}
\begin{tabular}{cccccccccccccc}
\toprule
& & & & & & & \multicolumn{3}{c}{Coverage} & \multicolumn{3}{c}{CI Length} \\
\cmidrule(lr){8-10} \cmidrule(lr){11-13}
$N$ & Estimator & Bias & RMSE & RMSE/SEB & PSE & PSE/SEB & Wald & SS & Boot & Wald & SS & Boot \\
\midrule
\multirow{9}{*}{1000} & IPW & 4.452 & 7.788 & 1.158 & - & - & - & - & 0.892 & - & - & 24.940 \\
 & CBPS & 4.454 & 7.789 & 1.158 & - & - & - & - & 0.891 & - & - & 24.937 \\
 & HAL-IPW & 1.482 & 6.913 & 1.029 & - & - & 0.941 & - & - & 26.625 & - & - \\ 
 & AIPW & 0.827 & 6.865 & 1.022 & - & - & 0.950 & - & - & 26.652 & - & - \\ 
 & MMD-G & 1.127 & 7.064 & 1.050 & - & - & - & 0.942 & 0.928 & - & 27.606 & 25.145 \\
 & MMD-EB & 1.486 & 7.112 & 1.057 & - & - & - & 0.945 & 0.877 & - & 27.342 & 21.711 \\
 & MMD-SKB & 4.395 & 7.815 & 1.162 & - & - & - & - & 0.899 & - & - & 25.241 \\
 & SL1DB & 1.336 & 7.106 & 1.057 & 7.040 & 1.047 & 0.952 & 0.950 & 0.893 & 27.598 & 27.886 & 22.383 \\
 & SL2DB & 1.395 & 7.093 & 1.055 & 7.037 & 1.046 & 0.953 & 0.940 & 0.895 & 27.585 & 27.719 & 22.177 \\ \midrule
\multirow{9}{*}{2000} & IPW & 4.098 & 6.200 & 1.302 & - & - & - & - & 0.833 & - & - & 17.599 \\
 & CBPS & 4.098 & 6.200 & 1.302 & - & - & - & - & 0.833 & - & - & 17.598 \\
 & HAL-IPW & 0.901 & 4.837 & 1.017 & - & - & 0.949 & - & - & 18.637 & - & - \\ 
 & AIPW & 0.516 & 4.821 & 1.014 & - & - & 0.941 & - & - & 18.551 & - & - \\ 
 & MMD-G & 0.402 & 4.929 & 1.035 & - & - & - & 0.949 & 0.934 & - & 19.723 & 17.967 \\
 & MMD-EB & 0.682 & 4.981 & 1.046 & - & - & - & 0.949 & 0.883 & - & 19.364 & 15.437 \\
 & MMD-SKB & 4.104 & 6.220 & 1.306 & - & - & - & - & 0.839 & - & - & 17.674 \\
 & SL1DB & 0.545 & 4.994 & 1.049 & 5.004 & 1.051 & 0.959 & 0.948 & 0.891 & 19.614 & 19.845 & 15.843 \\
 & SL2DB & 0.570 & 4.994 & 1.049 & 4.990 & 1.048 & 0.956 & 0.946 & 0.892 & 19.561 & 19.754 & 15.732 \\ \midrule
\multirow{9}{*}{4000} & IPW & 4.252 & 5.321 & 1.581 & - & - & - & - & 0.734 & - & - & 12.442 \\
 & CBPS & 4.252 & 5.322 & 1.581 & - & - & - & - & 0.735 & - & - & 12.442 \\
 & HAL-IPW & 0.430 & 3.440 & 1.022 & - & - & 0.946 & - & - & 13.184 & - & - \\ 
 & AIPW & 0.215 & 3.426 & 1.018 & - & - & 0.945 & - & - & 13.029 & - & -  \\ 
 & MMD-G & 0.301 & 3.397 & 1.009 & - & - & - & 0.945 & 0.935 & - & 14.078 & 12.831 \\
 & MMD-EB & 0.464 & 3.480 & 1.034 & - & - & - & 0.943 & 0.882 & - & 13.898 & 10.966 \\
 & MMD-SKB & 0.791 & 3.428 & 1.018 & - & - & - & - & 0.938 & - & - & 12.882 \\
 & SL1DB & 0.370 & 3.464 & 1.029 & 3.536 & 1.050 & 0.946 & 0.941 & 0.892 & 13.859 & 14.165 & 11.217 \\
 & SL2DB & 0.389 & 3.457 & 1.027 & 3.529 & 1.049 & 0.947 & 0.942 & 0.896 & 13.834 & 14.090 & 11.173 \\
\bottomrule
\end{tabular}
\caption{\footnotesize Point estimation and inference performance. Bias, RMSE, PSE, and CI lengths are scaled by a factor of 100.}
\label{tab:sim_point_estimation}
\end{table}

Finally, Table~\ref{tab:sim_time} highlights the computational advantages of \method. HAL-IPW and AIPW are by far the most expensive methods for point estimation at every sample size, with their Wald intervals adding negligible cost. \method is also faster than MMD-G and MMD-EB for point estimation, with larger gains under subsampling, and its Wald interval is computed essentially instantly from \eqref{eq-plugin variance-c}. Overall, the results reveal a clear trade-off: the faster methods (IPW, CBPS, and MMD-SKB) are either biased or lack theoretical guarantees, whereas those matching \method's statistical performance (HAL-IPW, AIPW, MMD-G, and MMD-EB) are considerably more expensive, and HAL-IPW and AIPW further use outcome information. \method thus combines small bias, near-efficient RMSE, and valid inference with a closed-form interval, without using outcome information and at a moderate computational cost, at least at the sample sizes considered.

\begin{table}[!htp]
\centering
\scriptsize
\setlength{\tabcolsep}{3pt}
\begin{tabular}{ccccccccccccc}
\toprule
$N$          & \multicolumn{4}{c}{1000}                                                                                               & \multicolumn{4}{c}{2000}                                                                                 & \multicolumn{4}{c}{4000}                                                                                                 \\
Estimator    & Point Est.                    & Wald & SS                              & Boot                            &  Point Est.                    & Wald & SS                              & Boot & Point Est.                    & Wald & SS                              & Boot                    \\
\cmidrule(lr){1-1} \cmidrule(lr){2-5} \cmidrule(lr){6-9} \cmidrule(lr){10-13}
IPW          & \phantom{11}0.08 & -    & -                                & \phantom{11}7.63 & \phantom{111}0.08 & -    & -     & \phantom{11}11.23 & \phantom{11}0.09 & -    & -                                  & \phantom{111}18.12 \\
CBPS         & \phantom{11}0.11 & -    & -                                & \phantom{1}10.25 & \phantom{111}0.12 & -    & -     & \phantom{11}15.25 & \phantom{11}0.15 & -    & -                                  & \phantom{111}24.76 \\
HAL-IPW      & 492.39          & 0.00 & -                                & -                & 1073.26           & 0.00 & -     & -                 & 832.46           & 0.00 & -                                  & -                  \\
AIPW         & 139.65           & 0.00 & -                                & -                & \phantom{1}284.58 & 0.00 & -     & -                 & 625.87           & 0.00 & -                                  & -                  \\
MMD-G        & \phantom{11}1.27 & -    & 59.02                            & 404.63           & \phantom{111}9.25 & -    & 87.78 & 3424.23           & \phantom{1}62.85 & -    & 152.47                             & 24167.55           \\
MMD-EB       & \phantom{11}1.21 & -    & 55.14                            & 418.20           & \phantom{111}8.92 & -    & 84.57 & 3374.44           & \phantom{1}63.17 & -    & 151.78                             & 24255.17           \\
MMD-SKB      & \phantom{11}0.17 & -    & -                                & \phantom{1}13.45 & \phantom{111}0.18 & -    & -     & \phantom{11}20.37 & \phantom{11}0.55 & -    & -                                  & \phantom{11}174.35 \\
SL1DB        & \phantom{11}0.73 & 0.04 & 15.23                            & 290.55           & \phantom{111}5.19 & 0.05 & 24.65 & 2606.24           & \phantom{1}29.80 & 0.05 & \phantom{1}55.86                   & 17327.03           \\
SL2DB        & \phantom{11}0.86 & 0.04 & \phantom{1}6.87                  & 319.45           & \phantom{111}4.37 & 0.05 & 14.61 & 2013.02           & \phantom{1}24.11 & 0.05 & \phantom{1}45.69                   & 11482.69           \\
\bottomrule
\end{tabular}

\caption{\footnotesize Average computation time in seconds. For subsampling-based CIs, computation time includes the data-driven selection of $m$ described in Appendix A.}
\label{tab:sim_time}
\end{table}

\section{Application} \label{sec:application}
In this section, we apply the proposed \method method to evaluate the causal effect of 401(k) participation on net financial assets. The dataset is sourced from the empirical study by~\cite{Chernozhukov2018}. Among $n=\ $9,915 individuals included, 3,682 of them are eligible for a 401(k) plan.
Our primary objective is to understand the causal impact of the provision of the 401(k) plan on net assets. Since we are specifically interested in the effect on those who are induced to participate by being eligible for the plan, we employ an instrumental variable approach to estimate the LATE. Detailed discussions regarding the identification strategy are provided in Appendix C.

The outcome variable is the net total financial assets, the treatment variable is an indicator of whether an individual participates in a 401(k) plan, and the instrumental variable is an indicator of whether the employer offers a 401(k) plan. Additionally, we adjust for nine pre-treatment covariates, which are age, income, family size, years of education, defined benefit pension status, marital status, two-earner household status, individual retirement account (IRA) participation, and homeownership. All covariates are rescaled prior to the analysis.

We apply the competing estimators evaluated in Section~\ref{sec:simulation} to the 401(k) dataset ($N = 9{,}915$), with implementation and hyperparameter tuning configurations detailed in Appendix A. The empirical results are summarised in Table~\ref{tab:application}.

For point estimation, the parametric methods (IPW and CBPS) produce highly unstable LATE estimates and wide confidence intervals that cover zero. This instability reflects the sensitivity of parametric propensity score models to potential model misspecification in complex observational settings. In contrast, all nonparametric methods (HAL-IPW and AIPW) and distributional balancing methods (Gaussian, EB, SKB, SL1DB, and SL2DB) robustly identify a statistically significant, positive LATE of 401(k) participation on net financial assets, with point estimates clustering between $\$11{,}100.0$ and $\$12{,}400.0$.

For inference, both \method estimators produce narrower confidence intervals than MMD-G and MMD-EB. While MMD-SKB yields an interval length comparable to that of \method under subsampling, we emphasise that MMD-SKB relies on a low-rank approximation to the Gram matrix and lacks formal theoretical guarantees for statistical inference. HAL-IPW and AIPW yield the shortest intervals, about $15\%$ shorter than the Wald intervals of SL1DB and SL2DB, which we see as an instance of the conservativeness of our plug-in variance estimator; yet, subsampling confidence intervals of \method are comparable to those of HAL-IPW and AIPW. 

Finally, the results highlight the computational advantages of \method. MMD-G and MMD-EB rely on dense quadratic programming with $\mathcal{O}(n^3)$ complexity, whereas \method avoids this bottleneck via one-dimensional sorting, yielding substantial speedups in both point estimation and subsampling, and a valid Wald interval at minimal cost. HAL-IPW and AIPW are also considerably more expensive than \method for point estimation, although much faster here than in the simulations, possibly because most covariates in this dataset are binary or multicategory, which simplifies the optimisation path of their learners.

\begin{table}[htbp]
\centering
\scriptsize
\setlength{\tabcolsep}{2.5pt}
\begin{tabular}{ccccccc}
\toprule
Method & Inference & Estimate & 95\% CI & Length & Time (Point Est.) & Time (CI) \\
\midrule
IPW & Boot & \phantom{1}2,427.6 & [$-10{,}185.8$, $11{,}304.1$] & 21,489.9 
& \phantom{11}0.04 & \phantom{1}11.61 \\
CBPS & Boot & \phantom{1}5,769.1 & \phantom{11\,\,}[$-961.7$, $11{,}881.3$] & 12,843.0 
& \phantom{11}0.09 & \phantom{1}18.95 \\
HAL-IPW & Wald & 11,692.6 & \phantom{11\,\,}[$8{,}527.4$, $14{,}857.8$] & \phantom{1}6,330.4 & \phantom{1}67.67 & \phantom{11}0.00 \\
AIPW & Wald & 11,845.6 & \phantom{11\,\,}[$8{,}683.6$, $15{,}007.5$] & \phantom{1}6,323.9 & 209.39 & \phantom{11}0.00 \\
MMD-Gaussian & SS & 11,877.6 & \phantom{11\,\,}[$6{,}608.0$, $16{,}171.8$] & \phantom{1}9,563.8 
& 205.82 & 254.39 \\
MMD-EB & SS & 11,943.4 & \phantom{11\,\,}[$8{,}038.2$, $15{,}778.0$] & \phantom{1}7,739.8 
& 208.27 & 265.77 \\
MMD-SKB & Boot & 12,342.6 & \phantom{11\,\,}[$9{,}025.2$, $15{,}546.3$] & \phantom{1}6,521.1 
& \phantom{11}1.11 & 550.22 \\
SL1DB & Wald & 11,630.5 & \phantom{11\,\,}[$7{,}924.5$, $15{,}336.5$] & \phantom{1}7,411.9 
& \phantom{1}16.99 & \phantom{11}0.01 \\
SL1DB & SS & 11,630.5 & \phantom{11\,\,}[$8{,}205.2$, $14{,}650.3$] & \phantom{1}6,445.2 
& \phantom{1}16.99 & \phantom{1}64.30 \\
SL2DB & Wald & 11,117.0 & \phantom{11\,\,}[$7{,}412.7$, $14{,}821.3$] & \phantom{1}7,408.7 
& \phantom{1}15.02 & \phantom{11}0.01 \\
SL2DB & SS & 11,117.0 & \phantom{11\,\,}[$7{,}773.6$, $14{,}263.0$] & \phantom{1}6,489.5 
& \phantom{1}15.02 & \phantom{1}51.50 \\
\bottomrule
\end{tabular}
\caption{\footnotesize Results of 401(k) Data Analysis ($N = 9{,}915$). Point estimates and CIs are reported in US Dollars (USD), and computation times are reported in seconds. For subsampling-based CIs, computation time includes the data-driven selection of $m$ described in Appendix A.}
\label{tab:application}
\end{table}

\section{Discussion}\label{sec:discussion}

In this paper, we introduced \method, a distributional balancing method based on non-RKHS IPMs. The non-RKHS structure of the underlying discrepancy poses distinct theoretical and computational challenges, necessitating careful development on both aspects. From a theoretical perspective, we established $\sqrt{n}$-consistency of the ATE estimator, a guarantee that is generally unavailable for distributional balancing methods based on non-RKHS IPMs. Furthermore, we established asymptotic normality of the ATE estimator without outcome regression augmentation, providing, to the best of our knowledge, the first such result for a distributional balancing estimator. From a computational perspective, our simulations and real-data applications demonstrated substantial computational gains over existing MMD-based distributional balancing methods while maintaining competitive statistical performance. 

Beyond estimation of the ATE, the \method framework can be extended to a broad range of causal estimands and settings, including the ATT, LATE, subgroup effects, multi-category treatments, and optimal treatment regimes. These methodological extensions are presented in Appendix C. Although not pursued in detail in this paper, the same distributional balancing principle can also be extended beyond causal inference, including estimation of population means under missing at random and survey sampling settings.

The \method framework developed here is specific to binary treatments, where balance is defined as alignment between two treatment-specific covariate distributions and the marginal distribution. Many applications instead involve continuous treatments (e.g., dosage) or dynamic treatment regimes, where this two-sample notion of balance does not directly apply. A promising direction for extending our framework to continuous treatments is to pursue independence weights. Rather than balancing treatment-specific covariate distributions against the marginal distribution, this approach seeks weights $w$ such that the weighted study population satisfies, asymptotically, the independence of the treatment $A$ and covariates $X$ \citep{Huling2024independence}. Extending the sliced mechanism to such settings requires reformulating the balancing condition itself, which we leave as a topic for future work.

\section*{Acknowledgments}
Conflicts of interest: None declared.

\section*{Funding}
This work was supported by the National Science Foundation grant numbers DMS-2515262 and DMS-2515263.

\section*{Data Availability}
The 401(k) dataset analyzed in this paper is publicly available via the \texttt{DoubleML} R package. Code for implementing the proposed method and replicating all numerical experiments is available at the GitHub repository \url{https://github.com/haoran32/Sliced-L-p-Distributional-Balancing}.

\bibliographystyle{apa}
\bibliography{references}

@article{moonboot,
  author        = {Dalitz, Christoph and Lögler, Felix},
  title         = {moonboot: An {R} Package Implementing m-out-of-n Bootstrap Methods},
  journal       = {The R Journal},
  year          = {2025},
  volume        = {17},
  number        = {3},
  pages         = {125--137},
  doi           = {10.32614/RJ-2025-031},
  issn          = {2073-4859},
  note          = {https://github.com/cdalitz/moonboot/},
}

@article{de2025data,
  author        = {De, Simion and Huling, Jared D},
  title         = {Data adaptive covariate balancing for causal effect estimation for high dimensional data},
  journal       = {arXiv:2512.18069},
  year          = {2025},
}

@article{cramer1936some,
  author        = {Cram{\'e}r, Harald and Wold, Herman},
  title         = {Some theorems on distribution functions},
  journal       = {Journal of the London Mathematical Society},
  year          = {1936},
  volume        = {1},
  number        = {4},
  pages         = {290--294},
  publisher     = {Wiley Online Library},
}

@article{athey2018approximate,
  author        = {Athey, Susan and Imbens, Guido W and Wager, Stefan},
  title         = {Approximate residual balancing: debiased inference of average treatment effects in high dimensions},
  journal       = {Journal of the Royal Statistical Society Series B: Statistical Methodology},
  year          = {2018},
  volume        = {80},
  number        = {4},
  pages         = {597--623},
  publisher     = {Oxford University Press},
}

@article{Zhao2019,
  author        = {Zhao, Qingyuan},
  title         = {{Covariate balancing propensity score by tailored loss functions}},
  journal       = {The Annals of Statistics},
  year          = {2019},
  volume        = {47},
  number        = {2},
  pages         = {965--993},
  publisher     = {Institute of Mathematical Statistics},
  doi           = {10.1214/18-AOS1698},
  url           = {https://doi.org/10.1214/18-AOS1698},
}

@article{ImaiRatkovic2014,
  author        = {Imai, K. and Ratkovic, M.},
  title         = {Covariate balancing propensity score},
  journal       = {Journal of the Royal Statistical Society: Series B},
  year          = {2014},
  volume        = {76},
  number        = {1},
  pages         = {243--263},
}

@article{Hainmueller2012,
  author        = {Hainmueller, J.},
  title         = {Entropy balancing for causal effects: {A} multivariate reweighting method to produce balanced samples in observational studies},
  journal       = {Political Analysis},
  year          = {2012},
  volume        = {20},
  number        = {1},
  pages         = {25--46},
}

@article{HulingMak2024,
  author        = {Huling, J. D. and Mak, S.},
  title         = {Energy balancing of covariate distributions},
  journal       = {Journal of Causal Inference},
  year          = {2024},
  volume        = {12},
  number        = {1},
}

@article{WongChan2018,
  author        = {Wong, R. K. W. and Chan, K. C. G.},
  title         = {Kernel-based covariate functional balancing for observational studies},
  journal       = {Biometrika},
  year          = {2018},
  volume        = {105},
  number        = {1},
  pages         = {199--213},
}

@article{ChanYamZhang2016,
  author        = {Chan, K. C. G. and Yam, S. C. P. and Zhang, Z.},
  title         = {Globally efficient non-parametric inference of average treatment effects by empirical balancing calibration weighting},
  journal       = {Journal of the Royal Statistical Society: Series B},
  year          = {2016},
  volume        = {78},
  number        = {3},
  pages         = {673--700},
}

@article{HorvitzThompson1952,
  author        = {Horvitz, D. G. and Thompson, D. J.},
  title         = {A generalization of sampling without replacement from a finite universe},
  journal       = {Journal of the American Statistical Association},
  year          = {1952},
  volume        = {47},
  number        = {260},
  pages         = {663--685},
}

@article{Rubin1974,
  author        = {Rubin, D. B.},
  title         = {Estimating causal effects of treatments in randomized and nonrandomized studies.},
  journal       = {Journal of Educational Psychology},
  year          = {1974},
  volume        = {66},
  number        = {5},
  pages         = {688--701},
  publisher     = {American Psychological Association},
}

@article{Rosenbaum1983,
  author        = {Rosenbaum, P. R. and Rubin, D. B.},
  title         = {The central role of the propensity score in observational studies for causal effects},
  journal       = {Biometrika},
  year          = {1983},
  volume        = {70},
  number        = {1},
  pages         = {41--55},
  publisher     = {Oxford University Press},
}

@inproceedings{Rabin2011,
  author        = {Rabin, J. and Peyr{\'e}, G. and Delon, J. and Bernot, M.},
  title         = {Wasserstein Barycenter and Its Application to Texture Mixing},
  booktitle     = {International Conference on Scale Space and Variational Methods in Computer Vision},
  year          = {2011},
  pages         = {435--446},
  organization  = {Springer},
}

@article{Bonneel2015,
  author        = {Bonneel, N. and Rabin, J. and Peyr{\'e}, G. and Pfister, H.},
  title         = {Sliced and Radon Wasserstein barycenters of measures},
  journal       = {Journal of Mathematical Imaging and Vision},
  year          = {2015},
  volume        = {51},
  number        = {1},
  pages         = {22--45},
  publisher     = {Springer},
}

@incollection{Robins2008,
  author        = {Robins, James M. and Li, Lingling and {Tchetgen Tchetgen}, Eric and van der Vaart, Aad W.},
  title         = {Higher order influence functions and minimax estimation of nonlinear functionals},
  booktitle     = {Probability and Statistics: Essays in Honor of {David A. Freedman}},
  year          = {2008},
  volume        = {2},
  pages         = {335--421},
  publisher     = {Institute of Mathematical Statistics},
  address       = {Beachwood, Ohio},
  series        = {Institute of Mathematical Statistics Collections},
  editor        = {Speed, Terry and Nolan, Deborah},
}

@book{Imbens2015,
  author        = {Imbens, Guido W. and Rubin, Donald B.},
  title         = {Causal Inference for Statistics, Social, and Biomedical Sciences: An Introduction},
  year          = {2015},
  publisher     = {Cambridge University Press},
  address       = {Cambridge},
  doi           = {10.1017/CBO9781139025751},
  isbn          = {9780521885881},
}

@book{Hernan2020,
  author        = {Hern{\'a}n, Miguel A. and Robins, James M.},
  title         = {Causal Inference: What If},
  year          = {2020},
  publisher     = {Chapman \& Hall/CRC},
  address       = {Boca Raton},
  url           = {https://miguelhernan.org/whatifbook},
}

@phdthesis{Bonnotte2013,
  author        = {Bonnotte, Nicolas},
  title         = {Unidimensional and Evolution Methods for Optimal Transportation},
  year          = {2013},
  school        = {Université Paris-Sud},
}

@article{HiranoImbensRidder2003,
  author        = {Hirano, Keisuke and Imbens, Guido W. and Ridder, Geert},
  title         = {Efficient Estimation of Average Treatment Effects Using the Estimated Propensity Score},
  journal       = {Econometrica},
  year          = {2003},
  volume        = {71},
  number        = {4},
  pages         = {1161--1189},
  doi           = {10.1111/1468-0262.00442},
}

@article{ChenChenYu2023,
  author        = {Chen, Rui and Chen, Guanhua and Yu, Menggang},
  title         = {Entropy Balancing for Causal Generalization with Target Sample Summary Information},
  journal       = {Biometrics},
  year          = {2023},
  volume        = {79},
  number        = {4},
  pages         = {3179--3190},
  doi           = {10.1111/biom.13825},
}

@article{ChenHulingChenYu2024,
  author        = {Chen, Rui and Huling, Jared D. and Chen, Guanhua and Yu, Menggang},
  title         = {Robust Sample Weighting to Facilitate Individualized Treatment Rule Learning for a Target Population},
  journal       = {Biometrika},
  year          = {2024},
  volume        = {111},
  number        = {1},
  pages         = {309--329},
  doi           = {10.1093/biomet/asad038},
}

@inproceedings{Kong2023,
  author        = {Kong, Insung and Park, Yuha and Jung, Joonhyuk and Lee, Kwonsang and Kim, Yongdai},
  title         = {Covariate Balancing Using the Integral Probability Metric for Causal Inference},
  booktitle     = {Proceedings of the 40th International Conference on Machine Learning},
  year          = {2023},
  volume        = {202},
  pages         = {17430--17461},
  publisher     = {PMLR},
  series        = {Proceedings of Machine Learning Research},
}

@inproceedings{Yan2024,
  author        = {Yan, Yuguang and Zhou, Hao and Yang, Zeqin and Chen, Weilin and Cai, Ruichu and Hao, Zhifeng},
  title         = {Reducing Balancing Error for Causal Inference via Optimal Transport},
  booktitle     = {Proceedings of the 41st International Conference on Machine Learning},
  year          = {2024},
  volume        = {235},
  publisher     = {PMLR},
  series        = {Proceedings of Machine Learning Research},
}

@techreport{VanHandel2014,
  author        = {van Handel, Ramon},
  title         = {Probability in High Dimension},
  year          = {2014},
  number        = {APC 550},
  note          = {Lecture notes},
  institution   = {Princeton University},
}

@techreport{Boyd2003,
  author        = {Boyd, Stephen and Xiao, Lin and Mutapcic, Almir},
  title         = {Subgradient Methods},
  year          = {2003},
  note          = {Lecture notes for EE392o, Autumn 2003},
  institution   = {Stanford University},
}

@article{Kallus2020,
  author        = {Kallus, Nathan},
  title         = {Generalized Optimal Matching Methods for Causal Inference},
  journal       = {Journal of Machine Learning Research},
  year          = {2020},
  volume        = {21},
  number        = {62},
  pages         = {1--54},
  url           = {http://jmlr.org/papers/v21/19-120.html},
}

@book{Efron1994,
  author        = {Efron, Bradley and Tibshirani, Robert J.},
  title         = {An Introduction to the Bootstrap},
  year          = {1994},
  volume        = {57},
  publisher     = {Chapman \& Hall/CRC},
  address       = {New York},
  series        = {Monographs on Statistics and Applied Probability},
}

@article{Politis1994,
  author        = {Politis, Dimitris N. and Romano, Joseph P.},
  title         = {Large sample confidence regions based on subsamples under minimal assumptions},
  journal       = {The Annals of Statistics},
  year          = {1994},
  volume        = {22},
  number        = {4},
  pages         = {2031--2050},
  publisher     = {Institute of Mathematical Statistics},
}

@book{Politis1999,
  author        = {Politis, Dimitris N. and Romano, Joseph P. and Wolf, Michael},
  title         = {Subsampling},
  year          = {1999},
  publisher     = {Springer},
  address       = {New York},
  series        = {Springer Series in Statistics},
}

@article{Zubizarreta2015,
  author        = {Zubizarreta, Jos{\'e} R},
  title         = {Stable weights that balance covariates for estimation with incomplete outcome data},
  journal       = {Journal of the American Statistical Association},
  year          = {2015},
  volume        = {110},
  number        = {511},
  pages         = {910--922},
  publisher     = {Taylor \& Francis},
  doi           = {10.1080/01621459.2015.1023805},
}

@article{Kim2024,
  author        = {Kim, Kwangho and Niknam, Bijan A and Zubizarreta, Jos{\'e} R},
  title         = {Scalable kernel balancing weights in a nationwide observational study of hospital profit status and heart attack outcomes},
  journal       = {Biostatistics},
  year          = {2024},
  volume        = {25},
  number        = {3},
  pages         = {736--753},
  publisher     = {Oxford University Press},
  doi           = {10.1093/biostatistics/kxad032},
}

@article{Hirshberg2021,
  author        = {Hirshberg, David A. and Wager, Stefan},
  title         = {Augmented minimax linear estimation},
  journal       = {The Annals of Statistics},
  year          = {2021},
  volume        = {49},
  number        = {6},
  pages         = {3206--3227},
  publisher     = {Institute of Mathematical Statistics},
  doi           = {10.1214/21-AOS2080},
  url           = {https://doi.org/10.1214/21-AOS2080},
}

@article{Chernozhukov2018,
  author        = {Chernozhukov, Victor and Chetverikov, Denis and Demirer, Mert and Duflo, Esther and Hansen, Christian and Newey, Whitney and Robins, James},
  title         = {Double/debiased machine learning for treatment and structural parameters},
  journal       = {The Econometrics Journal},
  year          = {2018},
  volume        = {21},
  number        = {1},
  pages         = {C1--C68},
  publisher     = {Oxford University Press},
}

@article{Shen2025,
  author        = {Shen, Andy A. and Ben-Michael, Eli and Feller, Avi and Keele, Luke and Murray, Jared},
  title         = {Forest Kernel Balancing Weights: Outcome-Guided Features for Causal Inference},
  journal       = {Statistics in Medicine},
  year          = {2025},
  volume        = {45},
  number        = {20-22},
  pages         = {e70720},
}

@incollection{Hajek1971,
  author        = {H{\'a}jek, Jaroslav},
  title         = {Comment on ``{A}n {E}ssay on the {L}ogical {F}oundations of {S}urvey {S}ampling, {P}art {O}ne'' by {D}. {B}asu},
  booktitle     = {Foundations of Statistical Inference},
  year          = {1971},
  pages         = {236},
  publisher     = {Holt, Rinehart and Winston},
  address       = {Toronto--Montreal},
  editor        = {Godambe, Vidyadhar P. and Sprott, David A.},
}

@article{Deville1992,
  author        = {Deville, Jean-Claude and S{\"a}rndal, Carl-Erik},
  title         = {Calibration Estimators in Survey Sampling},
  journal       = {Journal of the American Statistical Association},
  year          = {1992},
  volume        = {87},
  number        = {418},
  pages         = {376--382},
  publisher     = {Taylor \& Francis},
  doi           = {10.1080/01621459.1992.10475217},
}

@article{Kim2026,
  author        = {Kim, Jae Kwang and Kwon, Yonghyun and Qiu, Yumou},
  title         = {Bregman projection for calibration estimation in {S}urvey {S}ampling},
  journal       = {arXiv:2603.20780},
  year          = {2026},
}

@article{Hazlett2020,
  author        = {Hazlett, Chad},
  title         = {Kernel Balancing: {A} Flexible Non-Parametric Weighting Procedure for Estimating Causal Effects},
  journal       = {Political Analysis},
  year          = {2020},
  volume        = {28},
  number        = {3},
  pages         = {317--342},
  publisher     = {Cambridge University Press},
  doi           = {10.1017/pan.2019.45},
}

@article{Gretton2012,
  author        = {Gretton, Arthur and Borgwardt, Karsten M. and Rasch, Malte J. and Sch{\"o}lkopf, Bernhard and Smola, Alexander},
  title         = {A Kernel Two-Sample Test},
  journal       = {Journal of Machine Learning Research},
  year          = {2012},
  volume        = {13},
  number        = {25},
  pages         = {723--773},
  url           = {http://jmlr.org/papers/v13/gretton12a.html},
}

@book{Nocedal2006,
  author        = {Nocedal, Jorge and Wright, Stephen J.},
  title         = {Numerical Optimization},
  year          = {2006},
  publisher     = {Springer},
  doi           = {10.1007/978-0-387-40065-5},
  edition       = {2},
}

@article{Wang2020,
  author        = {Wang, Yixin and Zubizarreta, Jos{\'e} R.},
  title         = {Minimal dispersion approximately balancing weights: Asymptotic properties and practical considerations},
  journal       = {Biometrika},
  year          = {2020},
  volume        = {107},
  number        = {1},
  pages         = {93--105},
  doi           = {10.1093/biomet/asz050},
}

@article{Huling2024independence,
  author        = {Huling, Jared D. and Greifer, Noah and Chen, Guanhua},
  title         = {Independence Weights for Causal Inference with Continuous Treatments},
  journal       = {Journal of the American Statistical Association},
  year          = {2024},
  volume        = {119},
  number        = {546},
  pages         = {1657--1670},
  publisher     = {Taylor \& Francis},
  doi           = {10.1080/01621459.2023.2213485},
  url           = {https://doi.org/10.1080/01621459.2023.2213485},
}

@article{Hahn1998,
  author        = {Hahn, Jinyong},
  title         = {On the role of the propensity score in efficient estimation of average treatment effects},
  journal       = {Econometrica},
  year          = {1998},
  volume        = {66},
  number        = {2},
  pages         = {315--331},
  publisher     = {JSTOR},
}

@inproceedings{Nadjahi2021,
  author        = {Nadjahi, Kimia and Durmus, Alain and Jacob, Pierre E. and Badeau, Roland and Simsekli, Umut},
  title         = {Fast Approximation of the Sliced-Wasserstein Distance Using Concentration of Random Projections},
  booktitle     = {Advances in Neural Information Processing Systems},
  year          = {2021},
  volume        = {34},
  pages         = {12411--12424},
}

@article{ED2004,
  author        = {Sz{\'e}kely, G{\'a}bor J and Rizzo, Maria L},
  title         = {Testing for equal distributions in high dimension},
  journal       = {InterStat},
  year          = {2004},
  volume        = {5},
  number        = {16.10},
  pages         = {1249--1272},
  publisher     = {Citeseer},
}

@article{CFD2026,
  author        = {Santra, Diptanil and Chen, Guanhua and Park, Chan},
  title         = {Distributional Balancing for Causal Inference: A Unified Framework via Characteristic Function Distance},
  journal       = {arXiv:2601.15449},
  year          = {2026},
}

@article{Ertefaie2023,
  author  = {Ertefaie, Ashkan and Hejazi, Nima S. and van der Laan, Mark J.},
  title   = {Nonparametric Inverse-Probability-Weighted Estimators Based on the Highly Adaptive Lasso},
  journal = {Biometrics},
  volume  = {79},
  number  = {2},
  pages   = {1029--1041},
  year    = {2023},
  doi     = {10.1111/biom.13719}
}

@article{Robins1994,
  author  = {Robins, James M. and Rotnitzky, Andrea and Zhao, Lue Ping},
  title   = {Estimation of regression coefficients when some regressors are not always observed},
  journal = {Journal of the American Statistical Association},
  volume  = {89},
  number  = {427},
  pages   = {846--866},
  year    = {1994},
  doi     = {10.1080/01621459.1994.10476818}
}

@book{villani2003topics,
title = {Topics in Optimal Transportation},
author = {Villani, C{\'e}dric},
series = {Graduate Studies in Mathematics},
volume = {58},
year = {2003},
publisher = {American Mathematical Society},
address = {Providence, RI},
isbn = {978-0-8218-3312-4}
}

\end{document}